\documentclass{pasj00}

\begin{document}
\SetRunningHead{T. Morokuma et al.}{FOCAS High-$z$ SN Spectroscopy}
\Received{2009/07/30}
\Accepted{2009/11/06}

\title{
Subaru FOCAS Spectroscopic Observations \\of High-Redshift Supernovae
\footnote{
Based on data collected at Subaru Telescope, 
which is operated by the National Astronomical Observatory of Japan. 
}
}
 \author{%
   Tomoki \textsc{Morokuma}\altaffilmark{1,2}, 
   Kouichi \textsc{Tokita}\altaffilmark{3}, 
   Christopher \textsc{Lidman}\altaffilmark{4}, 
   Mamoru \textsc{Doi}\altaffilmark{3,5}, 
   Naoki \textsc{Yasuda}\altaffilmark{5,6}, 
   Greg \textsc{Aldering}\altaffilmark{7}, 
   Rahman \textsc{Amanullah}\altaffilmark{7,8}, 
   Kyle \textsc{Barbary}\altaffilmark{7}, 
   Kyle \textsc{Dawson}\altaffilmark{9}, 
   Vitaliy \textsc{Fadeyev}\altaffilmark{10}, 
   Hannah K. \textsc{Fakhouri}\altaffilmark{7}, 
   Gerson \textsc{Goldhaber}\altaffilmark{7}, 
   Ariel \textsc{Goobar}\altaffilmark{8}, 
   Takashi \textsc{Hattori}\altaffilmark{11}, 
   Junji \textsc{Hayano}\altaffilmark{3}, 
   Isobel M. \textsc{Hook}\altaffilmark{12,13}, 
   D. Andrew \textsc{Howell}\altaffilmark{14,15}, 
   Hisanori \textsc{Furusawa}\altaffilmark{1}, 
   Yutaka \textsc{Ihara}\altaffilmark{2,3}, 
   Nobunari \textsc{Kashikawa}\altaffilmark{1}, 
   Rob A. \textsc{Knop}\altaffilmark{16}, 
   Kohki \textsc{Konishi}\altaffilmark{6}, 
   Joshua \textsc{Meyers}\altaffilmark{7}, 
   Takeshi \textsc{Oda}\altaffilmark{17}, 
   Reynald \textsc{Pain}\altaffilmark{18}, 
   Saul \textsc{Perlmutter}\altaffilmark{7}, 
   David \textsc{Rubin}\altaffilmark{7}, 
   Anthony L. \textsc{Spadafora}\altaffilmark{7}, 
   Nao \textsc{Suzuki}\altaffilmark{7}, 
   Naohiro \textsc{Takanashi}\altaffilmark{1}, 
   Tomonori \textsc{Totani}\altaffilmark{17}, 
   Hiroyuki \textsc{Utsunomiya}\altaffilmark{3}, 
   Lifan \textsc{Wang}\altaffilmark{19} \\
   (Supernova Cosmology Project)
 }
\altaffiltext{1}{National Astronomical Observatory of Japan, 2-21-1 Osawa, Mitaka, Tokyo 181-8588, Japan}
\altaffiltext{2}{Research Fellow of the Japan Society for the Promotion of Science}
\altaffiltext{3}{Institude of Astronomy, Graduate School of Science, University of Tokyo, 2-21-1, Osawa, Mitaka, Tokyo 181-0015, Japan}
\altaffiltext{4}{Oskar Klein Centre, Stockholm University, AlbaNova University Centre, 10691 Stockholm, Sweden}
\altaffiltext{5}{Institute of Physics and Mathematics of the Universe, University of Tokyo, Kashiwa, Chiba 277-8582, Japan}
\altaffiltext{6}{Institute for Cosmic Ray Research, University of Tokyo, Kashiwa, Chiba 277-8582, Japan}
\altaffiltext{7}{Lawrence Berkeley National Laboratory, 1 Cyclotron Road, Berkeley, CA 94720, USA}
\altaffiltext{8}{Physics Department, Stockholm University, AlbaNova University Centre, 10691 Stockholm, Sweden}
\altaffiltext{9}{Department of Physics and Astronomy, University of Utah, Salt Lake City, UT 84112, USA}
\altaffiltext{10}{Santa Cruz Institute for Particle Physics, University of Californica, Santa Cruz, CA 94064, USA}
\altaffiltext{11}{Subaru Telescope, National Astronomical Observatory of Japan, 650 North A'ohoku Place, Hilo, HI 96720, USA}
\altaffiltext{12}{University of Oxford Astrophysics, Denys Wilkinson Building, Keble Road, Oxford OX1 3RH, UK}
\altaffiltext{13}{INAF - Observatorio Astronomico di Roma, via Frascati 33, 00040, Monteporzio (RM), Italy}
\altaffiltext{14}{Las Cumbres Observatory Global Telescope Network, 6740 Cortona Dr., Suite 102, Goleta, CA 93117, USA}
\altaffiltext{15}{Department of Physics, University of California, Santa Barbara, Broida Hall, Mail Code 9530, Santa Barbara, CA 93106-9530}
\altaffiltext{16}{Department of Physics and Astronomy, Vanderbilt University, P.O. Box 1807, Nashville, TN 37240, USA}
\altaffiltext{17}{Department of Astronomy, Kyoto University, Sakyo-ku, Kyoto 606-8502, Japan}
\altaffiltext{18}{LPNHE, CNRS-IN2P3 and University of Paris VI and VII, 75005 Paris, France}
\altaffiltext{19}{Physics Department, Texas A\&M University, College Station, TX 77843, USA}
 \email{tomoki.morokuma@nao.ac.jp}
\KeyWords{stars: supernovae: general -- cosmology: observations-- surveys}

\maketitle

\begin{abstract}
  We present spectra of high-redshift supernovae (SNe) that were taken
  with the Subaru low resolution optical spectrograph, FOCAS.  These
  SNe were found in SN surveys with Suprime-Cam on Subaru, the CFH12k
  camera on the Canada-France-Hawaii Telescope (CFHT), 
  and the Advanced Camera for Surveys (ACS) on the Hubble Space Telescope
  (HST). These SN surveys specifically targeted $z>1$ Type~Ia
  supernovae (SNe~Ia). From the spectra of 39 candidates, we obtain
  redshifts for 32 candidates and spectroscopically identify 7 active 
  candidates as probable SNe~Ia, including one at $z=1.35$, which is the 
  most distant SN~Ia to be spectroscopically confirmed with a ground-based 
  telescope. 
  An additional 4 candidates are identified as likely
  SNe~Ia from the spectrophotometric properties of their host
  galaxies. Seven candidates are not SNe~Ia, either being SNe of
  another type or active galactic nuclei. 
  When SNe~Ia are observed within a week of maximum light, we find
  that we can spectroscopically identify most of them up to
  $z=1.1$. Beyond this redshift, very few candidates were
  spectroscopically identified as SNe~Ia. The current generation of
  super red-sensitive, fringe-free CCDs will push this redshift limit
  higher.
\end{abstract}

\section{Introduction}\label{sec:introduction}

Type~Ia supernovae (SNe~Ia) have proven to be very good standard 
candles for cosmological studies.  They are bright enough to detect at
cosmological distances, $z\sim1.5$, if one uses 8--10~m class ground-based
optical telescopes or the Hubble Space Telescope (HST), and their 
luminosities can be standardised using empirical relations between
luminosity and light curve shape (\cite{phillips1993})
and colour \citep{tripp1998}.

SNe~Ia have played a leading role in measuring the expansion history
of the universe since two independent teams, the Supernova Cosmology
Project (SCP) and the High-$z$ Team, discovered the accelerating
expansion of the universe (\cite{perlmutter1999}; \cite{riess1998}).
Since then, many projects to discover and identify SNe~Ia have been
organized. For example, the Carnegie Supernova Project (CSP;
\cite{hamuy2006}), the Nearby Supernova Factory (SNfactory;
\cite{aldering2002}), the Harvard-Smithsonian Center for Astrophysics
(CfA) supernova survey (\cite{jha2006}; \cite{hicken2009}), the Sloan
Digital Sky Survey-II Supernova Survey (SDSS-II SN Survey;
\cite{sako2008}; \cite{frieman2008}), the Supernova Legacy Survey
(SNLS; \cite{astier2006}), the Equation of State: SupErNovae trace
Cosmic Expansion (ESSENCE) survey
(\cite{miknaitis2007,woodvasey2007}), and Higher-Z team
(\cite{riess2007}) have detected around 1000 SNe up to redshift
$z\sim 1.5$.  The combination of SN~Ia data with measurements of
cosmic microwave background (CMB) fluctuations (\cite{spergel2003};
\cite{spergel2007}; \cite{komatsu2009}), baryon acoustic oscillations
(BAO; \cite{eisenstein2005}) and galaxy cluster number counts
(\cite{vikhlinin2009}) have constrained the dark energy equation of
state parameter.  However, the limits are consistent with very
different dark energy models, so the fundamental nature of dark energy
remains unclear. It is currently one of the biggest mysteries in
physics, and combined astronomical observations (SNe~Ia, CMB, BAO and 
weak lensing) seem to be the only way to constrain its properties.

Since the discovery of the accelerating expansion of the universe, the
SCP has been carrying out imaging surveys for SNe~Ia at $z\gtrsim 1$,
an epoch during which the expansion of the universe is expected to be
decelerating.  Spectroscopic follow-up observations are an essential
part of these surveys, providing spectroscopically determined
redshifts and, when necessary and possible, direct confirmation of the
SN type. In this paper, we present spectra of SNe and their host
galaxies taken with the Faint Object Camera And Spectrograph (FOCAS;
\cite{kashikawa2002}) on the Subaru 8.2-m telescope. Twelve SN
candidates shown here were found in ground-based observations
targeting blank fields and nearby galaxy clusters with Suprime-Cam
(\cite{miyazaki2002}) on Subaru and the CFH12k camera
(\cite{cuillandre2000}) on the Canada-France-Hawaii telescope (CFHT).
The remaining 27 SN candidates were discovered using the Advanced
Camera for Surveys (ACS; \cite{benitez2003}) on HST\footnote{Based on
  observations made with the NASA/ESA Hubble Space Telescope and
  obtained from the data archive at the Space Telescope
  Institute. STScI is operated by the Association of Universities for
  Research in Astronomy, Inc. under the NASA contract NAS 5-26555.
  The observations are associated with program 10496.} targeting high
redshift galaxy clusters in a program called the HST Cluster SN Survey
(Program number 10496, PI: Perlmutter). These SN searches specifically
targeted $z>1$ SNe~Ia, for which there have been relatively few
spectroscopically confirmed SNe~Ia (\cite{aldering1998}; 
\cite{coil2000}; \cite{tonry2003}; \cite{barris2004}; 
\cite{riess2004}; \cite{lidman2005}; \cite{riess2007}). 
In \S\ref{sec:observationanddata}, we summarize the SN searches and
present data for both imaging and spectroscopy.  Spectroscopic data
reductions are described in \S\ref{sec:observationanddata}.  SN and
host galaxy classifications are shown in \S\ref{sec:sntyping} and
\S\ref{sec:focashostspectraforhstclustersn}, respectively.  In
\S\ref{sec:typeideff}, we describe factors that influence the
classification of SNe.  \S\ref{sec:summary} is a summary of the paper.
We use the standard $\Lambda$CDM cosmological parameters of $(H_0,
\Omega_M, \Omega_\Lambda)=(70, 0.27, 0.73)$ for calculating the age of
the universe at a certain redshift. All magnitudes are measured in the
AB system.

\section{Observations and Data}\label{sec:observationanddata}

The spectroscopic observations described in this paper were carried
out during our SN campaigns that were specifically targeting $z>1$ SNe
Ia.  Each campaign generally consisted of broad band imaging to
discover SNe~Ia, spectroscopic follow-up to determine SN types and
redshifts, and photometric follow-up to measure light curves.  To 
first find and then effectively follow $z>1$ SNe~Ia, we used 
both wide-field imagers on ground-based telescopes, namely Suprime-Cam 
($34'\times27'$) on the Subaru telescope and the CFH12k camera
($42'\times28'$) on CFHT, and ACS ($3.3'\times3.3'$) on HST. 
SN candidates shown in this paper were found during 2001, 2002, 2005 and 2006.

\subsection{SN Search Campaigns}\label{sec:sncandidatesearch}

Our search for SNe Ia was conducted in a series of four campaigns.
Three of these were conducted with ground based facilities
during the Northern Spring of 2001, the Northern Spring of 2002 and
the Northern Fall of 2002. The fourth was an ACS search conducted
during 2005 and 2006.  
Some campaigns consisted of multiple searches with several instruments. 
Details of the searches are given in Table~\ref{tab:tab_searchobservation}.
Additional details can be found in \cite{lidman2005}, 
\cite{yasuda2008}, and \cite{dawson2009}, and are briefly summarised here. 
Some of the SN discoveries were reported in IAU circulars 
(\cite{doi2001}; \cite{yasuda2002}; \cite{doi2003}; \cite{dawson2006}).

\begin{table*}
\begin{center}
  \caption{ Summary of the SN candidate search campaigns.  The
    campaign name and search fields are denoted in columns 1 and 2.
    The telescope and instrument used for the SN search, and the
    number of pointings are denoted in columns 3 and 4.  The search
    season, the broad-band filter used in the search, the number of
    search epochs, and the number of spectra shown in this paper are
    denoted in column 5, 6, 7, and 8, respectively. $N_{\rm{spec}}$
    for HST Cluster Supernova Survey is the number of SN candidates and host galaxies. }
\label{tab:tab_searchobservation}. 
\begin{tabular}{lllccccc}
\hline\hline
\multicolumn{1}{c}{Campaign} & Field & Telescope/Instrument & $N_{\rm{pointings}}$ & Year/Month & Band & $N_{\rm{epoch}}$ & $N_{\rm{spec}}$
\\\hline
Subaru 2001    & CL1604\_0,4        & Subaru/Suprime-Cam & 2  & 2001/04--2001/05 & $R_C$     & 2    & 1\\
Subaru 2001    & MS1520.1           & Subaru/Suprime-Cam & 1  & 2001/04--2001/05 & $i'$      & 2    & 1\\
Subaru 2001    & SDF                & Subaru/Suprime-Cam & 1  & 2001/04--2001/05 & $i'$      & 2    & 5\\
Spring 2002    & SDF                & Subaru/Suprime-Cam & 2  & 2002/04--2002/05 & $i'$      & 2    & 0\\
Spring 2002    & SDFe,SDFw          & Subaru/Suprime-Cam & 2  & 2002/04--2002/05 & $i'$      & 2    & 0\\
Spring 2002    & SS1,2,3,4          & Subaru/Suprime-Cam & 4  & 2002/03--2002/04 & $i'$      & 3    & 1\\
Spring 2002    & C02 fields         & CFHT/CFH12k        & 3  & 2002/03--2002/06 & $I$       & $\sim10$ & 0\\
Fall 2002      & SXDF               & Subaru/Suprime-Cam & 5  & 2002/09--2003/10 & $i'$      & 5--7 & 2\\
HST Cluster SN & galaxy clusters    & HST/ACS            & 25 & 2005/08--2006/08 & $z_{850}$ & 5--10& 8+5\\
\hline
\end{tabular}
\end{center}
\end{table*}

\subsubsection{The Subaru 2001 Campaign}\label{sec:subaruspringcampaigns}

The Subaru Suprime-Cam campaign conducted in the Spring of 2001 was a
pilot search for later SN searches with this instrument. The
relatively wide field-of-view and the high sensitivity provided by
Suprime-Cam enables the discovery of high-redshift SNe more
effectively than ACS \citep{yasuda2003}.  Only two- or three-epochs
were taken with Suprime-Cam, which means that these data are not, by
themselves, enough to derive light curves, although one candidate was
followed with WFPC2 on HST \citep{amanullah2009}. The search fields
were centered on two galaxy clusters, CL 1604+4321 ($z=0.90$;
\cite{lubin2004}) and MS 1520.1+3002 ($z=0.117$; \cite{stocke1991}),
and several blank fields, such as the Subaru Deep Field (SDF;
\cite{kashikawa2004}).  The SDF observations were carried out as a
part of the SDF project.  The imaging data for the SN searches were
obtained in either the $R_C$ or $i'$ filters.  Typical exposure times
and limiting magnitudes for each epoch were one hour and $R_C,
i'\sim26$ mag.  SNe found in these campaigns are named [field]-[SN~ID]
or [field][SN~ID].

The Suprime-Cam data are very useful for investigating SN rates even
though there are only 2 or 3 epochs.  By combining Suprime-Cam data in
Abell 2152 \citep{totani2005}, \citet{oda2008} obtained SN rates for
both SNe~Ia and core-collapse SNe, and, by using a theoretical model
that relates SNe rate to the star formation rate \citep{oda2005}, the
cosmic star formation history. This provides an estimate that is
independent of galaxy studies. It also takes advantage of the fact
that SNe can trace star formation activity even though galaxies are
either too diffuse or too faint to detect.  Using well-sampled light
curves from the SXDS data (\cite{morokuma2008} and \S.\ref{sec:sxdf}
of this paper), the cosmological evolution of the SN~Ia rate can be
measured (\cite{totani2008}; \cite{ihara2008}).

\subsubsection{The Spring 2002 Campaign}\label{sec:cfh12ksurveyin2002spring}

The Spring 2002 Campaign consisted of four searches; two searches with
Subaru, a CFHT search, and a CTIO search \citep{lidman2005}.  The two
Subaru searches used Suprime-Cam, were done back-to-back and searched
for SNe~Ia in the SDF field, two fields surrounding the SDF field
(SDFe and SDFw, which are to the east and west of the SDF,
respectively), and four other blank fields (SS1, SS2, SS3, and SS4).
The CFHT search was a rolling search, which was similar to the one
subsequently done by the SNLS \citep{astier2006} where each field was
observed 3 to 4 times per lunation over a period of several lunations.
These kinds of searches automatically provide well-sampled light
curves of discovered transients.  We did not follow CTIO candidates
with FOCAS. SNe discovered in this search are reported in
\citet{lidman2005}.

SN candidates found in this campaign were named S02-[SN~ID] and
C02-[SN~ID] for the Subaru and CFHT searches, respectively. 
Followup imaging observations of several high-redshift SNe were
obtained with HST/ACS and two ground based IR instruments: ISAAC
(Infrared Spectrometer And Array Camera; \cite{moorwood1998})
on the VLT and NIRI (Near InfraRed Imager and Spectrometer;
\cite{hodapp2003}) on Gemini-North. These observations will be
reported in Suzuki et al.~(in preparation)

\subsubsection{The Fall 2002 Campaign}\label{sec:sxdf}

In the Fall 2002 campaign, we took multi-epoch $i'$-band images of
five fields with Suprime-Cam (\cite{yasuda2008}).  The fields were
centered on the Subaru/XMM-Newton Deep Field (SXDF). The
Subaru/XMM-Newton Deep Survey (SXDS) project (\cite{sekiguchi2004};
\cite{sekiguchi2008}) has taken deep multi-wavelength data, from X-ray
to radio, of this region.

Compared to earlier Subaru campaigns, the search and follow-up of
candidates discovered during the Fall 2002 campaign was more
comprehensive.  Reference images taken during the end of September
2002 were compared with search images that were taken during the
beginning of November 2002. The typical exposure time was one hour
and the limiting magnitude was $i'\sim26$ mag.  The deep and
wide-field Suprime-Cam data provided us with $\sim100$ variable
objects over a timescale of 1 month (\cite{morokuma2008};
\cite{yasuda2008}).  Follow-up imaging observations of several
high-redshift SNe were carried out using HST/ACS in the optical and
NICMOS (Near Infrared Camera and Multi-Object Spectrometer;
\cite{thompson1999}) on HST in the near infrared.  We also obtained
ground-based near-infrared $J$-band imaging data of one candidate with
both CISCO (Cooled Infrared Spectrograph and Camera for OH-airglow
Suppressor (\cite{iwamuro2001}, \cite{motohara2002})) on the Subaru
telescope and VLT/ISAAC. 
These observations will be reported in Suzuki et al.~(in preparation). 
SN candidates found in this campaign are named SuF02-[SN~ID].

\subsubsection{The HST Cluster Supernova Survey}\label{sec:hstclustersn}

The HST Cluster Supernova Survey was conducted in 2005 and 2006 with
ACS on HST (\cite{dawson2009}).  The targeted fields were galaxy
clusters at $0.9<z<1.5$. Galaxy clusters are rich in early-type
galaxies, which are expected to have little dust.  In current SN~Ia
cosmological studies, the systematic uncertainty in the correction for
dust attenuation is comparable to statistical uncertainties. For
example, it is not clear if the extinction law is universal or
unevolving.  The most straightforward way to avoid (or reduce) this
systematic error is to find SNe~Ia in dust-free environments.  Also,
the dispersion in the corrected peak $B$-band luminosity of SNe~Ia in
early-type galaxies is smaller than that in late-type galaxies (e.g.,
\cite{takanashi2008}).  Therefore, we searched for SNe Ia in
early-type galaxies by targeting galaxy clusters, which is an
effective way of finding a large number of early type galaxies at high
redshift in small fields of view.  Another advantage in targeting
early-type galaxies, where star-forming activities have been quenched,
is that SNe found in such galaxies are expected to be SNe~Ia with high
probability, as there should be no core-collapse SNe in these
galaxies.

In total, 25 high redshift galaxy clusters were targeted.  Typical exposure
times and limiting magnitudes for each epoch were 30 minutes and
F850LP ($z_{850}$) $\sim25$ mag, respectively.  The clusters were also
observed using the ACS F775W ($i_{775}$) filter and active high
redshift SNe were observed with the F110W filter on NICMOS.  The SN
candidates found in this campaign are named SCP[year][Cluster ID][SN
ID] where ``year'' corresponds to the year of discovery, ``Cluster ID'' is
a letter arbitrarily assigned to each cluster for scheduling purposes,
and ``SN~ID'' is a number assigned to each SN within a cluster according
to its time of discovery in the search.

\subsection{FOCAS Follow-up Spectroscopic Observations}\label{sec:followupspectroscopicobservations}

Spectroscopic follow-up is used to obtain redshifts and, if the SN phase was
near maximum light, the SN type. The candidates observed with
FOCAS are summarized in Tables \ref{tab:obssummary1} and
\ref{tab:obssummary2}.  Spectroscopic follow-up was also done with 
other facilities and is reported elsewhere (\cite{lidman2005}; \cite{yasuda2008}).
During the Fall 2002 campaign, the Subaru telescope was used mainly for 
discovering SNe, so the number of SN candidates observed spectroscopically 
with FOCAS was small for this particular campaign.

We used the long slit mode of FOCAS until December 2005 and then
mainly the multi-object slit mode after that.  There are two reasons
for the change.  First, candidates from the HST Cluster
Supernova Survey campaign were faint, which increased the difficulty
in acquiring these targets with the long slit mode.  The multi-object
slitmasks can be manufactured to include several alignment holes to
catch bright stars within the FOCAS FOV (6' diameter), thus enabling
us to center targets more precisely.  Slitmask observations also allow
us to observe additional targets, such as cluster members, at the same
time (\cite{eisenhardt2008}; \cite{tanaka2008}; \cite{dawson2009};
\cite{meyers2009}).  Also, improvements in the mask alignment software
by the FOCAS instrument team allowed us to acquire targets more
quickly.  Slit widths were 0\farcs8 in most of the observations. When
the seeing was poor, slit widths of 1\farcs0 were used.  The
atmospheric dispersion corrector was used in all observations.
Combinations of grisms and order-sort filters are 300B and SY47, or
300R and SO58, which result in spectra covering the 4700-9000\AA\ and
5800-10000\AA\ regions, respectively.  
Each exposure lasted 20 to 30 minutes. 
The objects were dithered by several arcseconds 
over several positions along the slit in order to effectively 
remove bad pixels, cosmic rays, and detector fringes. 
Exposure times varied with target brightness and priority. 
Typically, the seeing was 0\farcs6 to 1\farcs0 and 
total exposure times were 1-3 hours. 

\begin{table*}
\begin{center}
  \caption{ Summary of FOCAS Observations. Column 1 denotes observing programs 
    for FOCAS observations and allocated nights in parentheses. 
    Column 2 denotes observing dates in yyyy/mm/dd.  Exposure times (in seconds), target names,
    spectroscopic modes, grism/order-sort filter configurations, and
    standard stars used for flux calibration are listed in columns 3,
    4, 5, 6, and 7.  The letters ``LS'' indicate that the long-slit mode
    was used. When the multi-object spectroscopic mode was used, the
    slitmask name is used instead.  The spectroscopic configuration
    was either 300B/SY47 or 300R/SO58, which provide spectra over
    the 4700-9000\AA\ and 5800-10000\AA\ wavelength ranges, respectively. 
    In addition to the night listed below, 6 nights were lost to clouds.
  }
\label{tab:obssummary1}
\scriptsize
\begin{tabular}{lcclccc}
\hline\hline
\multicolumn{1}{l}{Program} & {Date (UT)} & Exposure & Targets & Mode & Configuration & Standard Star\\\hline
S01A-079 & 2001/05/26                  & $600\times1$     & SDF1        & LS & 300B/SY47 & Feige34\\
(2)      &                             & $2100\times3$    & SDF5        & LS & 300B/SY47 & Feige34\\
         & 2001/05/27                  & $900\times1$     & CL1604\_0-1 & LS & 300B/SY47 & Feige34\\
         &                             & $2100\times2$    & 1520.1-2    & LS & 300B/SY47 & Feige34\\
         &                             & $2100\times3$    & SDF2        & LS & 300B/SY47 & Feige34\\
         &                             & $2100\times1$    & SDF4        & LS & 300B/SY47 & Feige34\\
         &                             & $900\times1$     & SDF6        & LS & 300B/SY47 & Feige34\\\hline
S02A-174 & 2002/04/17                  & $1800\times3$    & S02-032     & LS & 300B/SY47 & Feige34\\
(1)      &                             & $1500\times1$    & C02-005     & LS & 300B/SY47 & Feige34\\
         &                             & $1800\times1$    & C02-007     & LS & 300B/SY47 & Feige34\\\hline
S02B-I04 & 2002/11/13                  & $2400\times3$    & SuF02-012   & LS & 300R/SO58 & Feige110\\
(2)      &                             & $1800\times3$    & SuF02-061   & LS & 300R/SO58 & Feige110\\\hline
S05B-137       & 2005/09/26 & $1800\times5$     & SCP05D0 & LS & 300R/SO58 & LTT2415\\
(1.5)          &            & $1800\times3$     & SCP05P1 & LS & 300R/SO58 & LTT2415\\
               & 2005/09/27 & $1800\times2$     & SCP06N10& LS & 300R/SO58 & GD71\\
               & 2005/09/28 & $1800\times6$     & SCP05D6 & LS & 300R/SO58 & GD71\\\hline
S05B-137       & 2005/10/26 & $1800\times(4+2)$\footnotemark[$*$] & SCP05D6 & LS & 300R/SO58 & BD+28d4211\\
(1.5)          &            & $1800\times3$     & SCP05P9 & LS & 300R/SO58 & BD+28d4211\\
               & 2005/10/27 & $1800\times4$     & SCP05D6 & LS & 300R/SO58 & BD+28d4211\\
               &            & $1800\times4$     & SCP05P9 & LS & 300R/SO58 & BD+28d4211\\
               & 2005/10/28 & $1800\times2$     & SCP06X13& LS & 300R/SO58 & GD71\\\hline
S05B-137 (0.5) & 2005/12/27 & $1800\times4$     & SCP06X18 & LS\footnotemark[$\dagger$] & 300R/SO58 & Feige34\\\hline
S05B-137       & 2006/04/22 & $1200\times5$ & SCP06F3, SCP06F6, SCP06F8 & F\_mask2B & 300R/SO58 & Feige34\\
(3)            & 2006/04/23 & $1200\times5$ & SCP06X26, SCP06X27        & X\_mask1  & 300R/SO58 & Feige34\\
               &            & $1200\times5$ & SCP06H3                   & H\_mask2  & 300R/SO58 & Feige34\\
               &            & $1200\times9$ & SCP06G3, SCP06G4          & G\_mask1  & 300R/SO58 & Wolf1346\\
               &            & $1200\times2$ & SCP06F3, SCP06F6, SCP06F8 & F\_mask2B & 300R/SO58 & Hz44\\
               & 2006/04/24 & $1200\times2$ & SCP06X26, SCP06X27        & X\_mask1  & 300R/SO58 & Feige34\\
               &            & $1200\times10$ & SCP06G3, SCP06G4         & G\_mask1  & 300R/SO58 & Wolf1346\\\hline
S05B-137       & 2006/06/28 & $1200\times4$ & SCP06F6, SCP06F12 & F\_mask2B & 300B/SY47 & Wolf1346\\
(1.5)          &            & $1200\times2$ & SCP06L22          & LS        & 300B/SY47 & Hz44\\
               & 2006/06/29 & $1200\times8$ & SCP06A4           & A\_mask1  & 300R/SO58 & BD+28d4211\\
               &            & $1200\times9$ & SCP06K0, SCP06K18 & K\_mask1  & 300R/SO58 & Hz44\\\hline
S06B-085       & 2006/08/23 & $1200\times2$ & SCP06F6           & LS        & 300B/SY47 & Wolf1346\\
(1)            &            & $1200\times6$ & SCP06B3, SCP06B4       & B\_mask1B & 300R/SO58 & Wolf1346\\
               &            & $1200\times3$ & SCP06V6                & V\_mask1  & 300R/SO58 & G191B2B\\
               &            & $1200\times6$ & SCP06N32, SCP06N33     & N\_mask1  & 300R/SO58 & G191B2B\\\hline
S06B-085       & 2006/12/24 & $1200\times8$ & SCP05D0, SCP05D6       & D\_mask1  & 300R/SO58 & BD+28d4211\\
(1)            &            & $1200\times6$ & SCP05D0, SCP05D6       & D\_mask2  & 300R/SO58 & BD+28d4211\\
               &            & $1200\times8$ & SCP06X18, SCP06X26, SCP06X27 & X\_mask2  & 300R/SO58 & Feige34\\
               &            & $1200\times4$ & SCP06E12             & E\_mask3  & 300R/SO58 & Feige34\\\hline
S06B-085       & 2007/05/18 & $1200\times6$ & SCP06E12             & E\_mask4  & 300R/SO58 & Hz44\\
(2)            &            & $1200\times6$ & SCP06T1     & T\_mask1  & 300R/SO58 & BD+28d4211\\
               & 2007/05/19 & $1200\times7$ & SCP06E12         & E\_mask6 & 300R/SO58 & Hz44\\
               &            & $1200\times7$ & SCP06T1 & T\_mask2 & 300R/SO58 & BD+28d4211
\\\hline
\multicolumn{5}{@{}l@{}}{\hbox to 0pt{\parbox{180mm}{\footnotesize
\par\noindent
\footnotemark[$*$] SCP05P9 was observed between the two exposures positions 
\par\noindent
\footnotemark[$\dagger$] Slit positioning may have failed as there was no signal.
}\hss}}
\end{tabular}
\end{center}
\end{table*}

\begin{table*}
  \caption{
    Summary of FOCAS spectrum fitting results. Columns 1 and 2 are target names. 
    In column 2, we list the IAU name, if the SN was reported in an IAU circular. 
    Otherwise,
    this field is either left blank or is filled with the nicknames of the candidates. 
    Columns 3 and 4 are the redshifts from the host galaxies and SNe. 
    If we could classify a candidate as a SN, we report the name and epoch of the best-fitting local SN 
    in columns 5 and 6, respectively.  Epochs from the light curve fits are denoted in column 7. 
    The confidence index (C.I.) of the classification and the object type are listed in columns 8 and 9, respectively. 
    Notes on individual objects are in column 10. 
  }
\label{tab:obssummary2}
\scriptsize
\begin{tabular}{llrrlrrccl}
\hline\hline
\multicolumn{1}{c}{Name 1} & Name 2 & $z_{\rm{gal}}$ & $z_{\rm{SN}}$ & template & $t_{\rm{sp}}$ & $t_{\rm{LC}}$\footnotemark[$*$]
& C.I.\footnotemark[$\dagger$]  & Type\footnotemark[$\dagger$] & Notes
\\\hline
CL1604\_0-1 & SN~2001cq  & -     & 0.36 & SN~1998aq    & $-8$  & -     & 5 & Ia     & Si\emissiontype{II},S\emissiontype{II}.\\
1520.1-2    & SN~2001cw  & 0.94  & 0.95 & Hsiao Ia     & $-6$  & $-5$  & 3 & Ia*    & 4000\AA\ break.\\
SDF1        & SN~2001cs  & 0.431 & 0.42 & SN~1994D     & $-1$  & -     & 5 & Ia     & Ca\emissiontype{II} H and K. Sig.~host galaxy contamination. \\
SDF2        & SN~2001ct  & -     & ?    & -            & -     & -     & 1 & II?    & Blue continuum.\\
SDF4        & SN~2001cr  & -     & ?    & -            & -     & -     & 2 & ?      & -\\
SDF5        & SN~2001cv  & 1.039 & 1.02 &Nugent Ia     & $20$  & -     & 2 & ?      & [O\emissiontype{II}]\footnotemark[$\S$].\\
SDF6        & SN~2001cu  & 0.515 & 0.50 &Nugent Ib/c   & $-6$  & -     & 2 & ?      & Ca\emissiontype{II} H and K. Sig.~host galaxy contamination. \\
            &            &       & 0.50 & Hsiao Ia     & $-6$  &       &   &        & Sig.~host galaxy contamination. \\
\hline
S02-032     & SN~2002ff  & -     & 1.02 & Nugent Ia    & $0$   & 0     & 4 & Ia     & -\\
C02-005     & -          & 0.426 & -    & -            & -     & -     & 0 & SB/AGN & Starburst or AGN.\\
C02-007     & -          & 1.780 & -    & -            & -     & -     & 0 & AGN    & AGN.\\\hline
SuF02-012   & SN~2002lc  & -     & 1.22? &  -         & -  & $-1$  & 2 & ? & Sig.~host galaxy contamination.\\
SuF02-061   & -          & 1.085 &       & -          & -  &$\sim5$& 2 & ? & Sig.~host galaxy contamination. AGN?\\\hline
SCP06A4     & Aki        & 1.19  &  -   & -            & -     & 0     & 2 & ?  & Low signal-to-noise ratio.\\
SCP06B3     & Isabella  & 0.744  & -    & -            & -     & 13    & 2 & ?  & [O\emissiontype{II}], H$\beta$, [O\emissiontype{III}].\\
SCP06B4     & Michaela  & 1.117  & -    & -            & -     & 4     & 2 & ?  & [O\emissiontype{II}], [Ne\emissiontype{III}].\\
SCP05D0     & Frida     & 1.015  & -    & -            & -     & 16    & - & -  & Ca\emissiontype{II} H and K.\\
            &           &        & -    & -            & -     & 241   & - & -  & Ca\emissiontype{II} H and K. Table~\ref{tab:fittingresult_host}.\\
SCP05D6     & Maggie    & 1.315  & -    & -            & -     & $-8$  & 2 & ?  & Ca\emissiontype{II} H and K.\\
            &           &        & -    & -            & -     & 4     & 2 & ?  & Ca\emissiontype{II} H and K.\\
            &           &        & -    & -            & -     & 187   & - & -  & Ca\emissiontype{II} H and K. Table~\ref{tab:fittingresult_host}.\\
SCP06E12    & -         & -      & -    & -            & -     & $>300$     & - & -& Nearby galaxy at z=1.025.\\
            &           &        & -    & -            & -     & $>480$     & - & -& -\\
SCP06F3     & -         & 1.21?  & -    & -            & -     & 0     & 2 & ?     & An emission line at 6195\AA?\ AGN?\\
SCP06F6     & -         & -      & -    & -            & -     & $-25$ & 0 & non-Ia& Blue continuum. \cite{barbary2009}.\\
            &           &        & -    & -            & -     &  42   & 0 & -     & Weak continuum.\\
            &           &        & -    & -            & -     &  98   & 0 & -     & Very low signal-to-noise ratio.\\
SCP06F8     & Ayako     & 0.789? & -    & -            & -     &  40   & - & -     & Spectrum dominated by the neighbor.\\
SCP06F12    & Caleb      & -      & -    & -            & -     & 194  & - & -     & Almost no continuum.\\
SCP06G3     & Brian     & 0.962  & -    & -            & -     & 24    & - & -     & [O\emissiontype{II}]\footnotemark[$\S$].\\
SCP06G4     & Shaya     & 1.350  & 1.35 & Hsiao Ia     & $-1$  & $-5$  & 3 & Ia*   & Probable Ca\emissiontype{II} H and K. Table~\ref{tab:fittingresult_host}.\\
SCP06H3     & Elizabeth & 0.851  & 0.84 & Hsiao Ia     & 2     & 0     & 4 & Ia    & [O\emissiontype{II}],[O\emissiontype{III}].\\
SCP06K0     & Tomo      & 1.416  & -    & -            & -     & 68    & - & -     & 4000\AA\ break. Table~\ref{tab:fittingresult_host}.\\
SCP06K18    & -         & 1.411  & -    & -            & -     & 84    & - & -     & 4000\AA\ break. Table~\ref{tab:fittingresult_host}.\\
SCP06L22                         & -      & 1.369 & - & -    & -     & -     & 0 & AGN& broad Mg\emissiontype{II}.\\
SCP06N10\footnotemark[$\ddagger$] & -      & 0.203 & - & -    & -     & $>54$ & - & - & [O\emissiontype{III}],H$\alpha$. \\
SCP06N32    & -         & -      & -    & -            & -     & $<3$  & 2 & Ib/c? & -\\
SCP06N33    & Naima     & 1.189  & -    & -            & -     & 4     & 2 & ?     & Very low signal-to-noise ratio.\\
SCP05P1     & Gabe      & 0.926  & -    & -            & -     & 16    & - & - & [O\emissiontype{II}],[O\emissiontype{III}].\\
SCP05P9     & Lauren    & 0.821  & 0.81 & Hsiao Ia     & $-2$  & $-3$  & 3 & Ia* & Sig.~host galaxy contamination. [O\emissiontype{II}],[O\emissiontype{III}],H$\beta$.\\
SCP06T1  & Jane  & 1.112?& - & - & - & 450  & - & - & Probable [O\emissiontype{II}].\\
SCP06V6  & -     & 0.903 & - & - & - & -    & 0 &AGN& [O\emissiontype{II}],Fe\emissiontype{II},broad H$\beta$,[O\emissiontype{III}].\\
SCP06X13 & -     & 1.642 & - & - & - & -    & 0 &AGN& Broad Mg\emissiontype{II}.\\
SCP06X18 & -     & -     & - & - & - & 16   & 2 & ? & Almost no continuum.\\
         &       &       & - & - & - & 378  & - & - & Almost no continuum.\\
SCP06X26 & Joe   & 1.44? & - & - & - & 67   & - & - & A marginal emission line at 9100\AA.\\
         &       &       & - & - & - & 312  & - & - & -\\
SCP06X27 & Olivia& 0.435 & - & - & - & 38   & - & - & Na\emissiontype{I} D,H$\alpha$,[N\emissiontype{II}].\\
         &       &       & - & - & - & 208  & - & - & Na\emissiontype{I} D,H$\alpha$,[N\emissiontype{II}].\\
\hline
\multicolumn{8}{@{}l@{}}{\hbox to 0pt{\parbox{180mm}{\footnotesize
\par\noindent
\footnotemark[$*$] In rest frame days, except for 
SCP06F6, SCP06E12, SCP06F3, SCP06F8, SCP06F12, SCP06N32, SCP06T1, SCP06X18, and SCP06X26, 
which are in observer frame days.\\
\footnotemark[$\dagger$] Generally left as ``-'' for targets that were observed more 
than two rest frame weeks after maximum light, where the host galaxy was the primary target.\\
\footnotemark[$\ddagger$] No fringe corrections were applied. A 1\farcs0-width slit was used.\\
\footnotemark[$\S$] The single emission line is identified as [O\emissiontype{II}].
}\hss}}
\end{tabular}
\end{table*}

\subsection{Spectroscopic Data Reduction}\label{sec:spectroscopicdatareduction}

The FOCAS\footnote{For the Subaru 2001 Campaign a detector mosaic 
of two 2k$\times$4k 3-side buttable CCDs fabricated by SITe were used. 
For all other campaigns,  a mosaic of MIT/LL CCDs were used.}
spectroscopic data were reduced in a
mostly standard way using the Image Reduction and Analysis Facility
(IRAF\footnote{IRAF is distributed by the National Optical Astronomy
  Observatories, which are operated by the Association of Universities
  for Research in Astronomy, Inc., under cooperative agreement with
  the National Science Foundation.}). The bias was removed by fitting
a low order polynomial to the overscan region, which lies parallel to
the dispersion direction, and the flat-fielding was done with lamp
flats. The background was removed while simultaneously masking bad
pixels and bad columns, and correcting for detector fringing, except
for those cases in which there were too few frames to compute a fringe
correction. The flexure is small enough for one to use all the
data taken with the same configuration during a night to make fringe
correction images.  The 2d spectral data were then stacked and the
1d spectrum of the object extracted.  Wavelength calibration was done
using night-sky emission lines.  The flux scale was calibrated using
the standard star spectrum obtained closest in time.  BD+28d4211,
Feige34, Feige110, Hz44, GD71, G191B2B, LTT2415, and Wolf1346 were
used as flux standards.  Most of them are HST spectrophotometric
standard stars with well-measured flux densities.
We also removed strong telluric absorption features using the 
standard star data that was used for flux calibration; however, we note 
that these corrections are not perfect, and when presenting the spectra 
we mark the main regions that may be affected by telluric features.

Fringes, which can be significant at long wavelengths, were removed
from the sky background to increase the signal-to-noise ratio of
the spectra.  To remove these fringe patterns effectively, we
adopted an observational and data reduction strategy that is described
below.  The procedure is similar to that described in
\citet{lidman2005}, which was used for data obtained with FORS1 and
FORS2 on the VLT.  The difference between ours and \citet{lidman2005}
comes from the specific properties (distortion, pixel scale, spectral
resolution, and so on) of each instrument.

All observations were dithered by several arcseconds along the
spatial direction of the slits, which are set perpendicular to
the dispersion direction.  In long slit spectroscopy, we use all the
exposures with the same configuration during a given night to create
a fringe frame for that set up and that night.  We also carefully designed
slitmasks for SCP05D0 and SCP05D6 (D\_mask1 and D\_mask2), SCP05E12
(E\_mask4 and E\_mask6), and SCP05T1 (T\_mask1 and T\_mask2) so that
the slit positions of these targets on the CCD were the same.
Data were sorted with respect to grism, slit width, date and the
location of the candidate on the 2D spectrum.  Fringe frames were
created by clipping deviant pixels (including pixels with flux
  from objects) and by averaging the reminder.  Since the intensity
of rows (the spatial axis of the spectra are along detector rows) can
vary with respect to one another, each row is treated individually.
The fringe frames were subtracted from the data after suitable
scaling.  Again, each row was treated individually. To help locate
cosmic rays, an average sky spectrum was added back to the 2
dimensional fringe corrected data. These data were then registered 
and combined, after clipping for cosmic rays, and a spectrum of the
candidate extracted.

\subsection{Spectral Fitting Method and SN Type Classification}\label{sec:spectrumfittingmethod}

We determine SN types by fitting spectra with spectral templates using
the software developed by \citet{tokita2009}. The software fits
  the spectra over a grid of parameters: redshift, supernova template,
  extinction, and, if necessary, the host galaxy template and the
  fraction of host galaxy light.  The templates consist of observed
spectra of all SN types available from the literature, the FOCAS
spectra of SDSS SNe (\cite{zheng2008}, \cite{tokita2009},
\cite{konishi2008}), and synthetic spectra of SN~Ia \citep{hsiao2007}
and the other non-Ia types \citep{nugent2002}.  The software computes
the reduced $\chi^{2}$ and determines the best combination of
type, redshift, extinction, and epoch of each SN for which the reduced
$\chi^{2}$ value is smallest.  The extinction law used in this paper
is from \cite{cardelli1989}.  We also consider host galaxy
contamination in some cases.  Errors in the template spectrum are
neglected.

When the redshift of the host, $z_{\rm{gal}}$, can be determined from
host galaxy spectral features, we allow the redshift of the SN,
$z_{\rm{SN}}$, to be fitted in the range
$\Delta{z}\equiv|(z_{\rm{gal}}-z_{\rm{SN}})|\sim\pm0.01-0.02$. The
range allows for velocity variations found in normal SNe and, to a
much smaller degree, a velocity offset due to the motion of the
SN with respect to the host.  When the light curve can be well fitted
by light curve templates, differences between the epochs, $\Delta{t}$,
are expected to be within $\sim\pm3$ days.  We note that epochs of
spectra $t$ and $\tau$ are defined as days from maximum light and
explosion, respectively, and that $t$ is generally applied for SNe~Ia
while $\tau$ is applied for core-collapse SNe.  Larger differences of
epochs are possible for later phases because the number of SN spectral
templates that are available is not so large.  For most of SN
candidates discovered during the Subaru 2001 and Spring 2002
campaigns, there are too few light curve epochs to set strong
constraints on the spectroscopic epoch.

In some cases, the extracted SN spectrum is contaminated by host
galaxy light. If the contamination is significant, we remove the
host galaxy light using one of two methods. In the case that the SN
and the host galaxy are well resolved, we subtract a spectrum that is
extracted from a region that is located on the other side of the
center of the host. We assume axial symmetry of the host galaxy.  When
the SN is located close to the center of the host galaxy, we subtract
a synthetic spectrum computed from the spectral library of
\cite{bruzual2003} (hereafter BC03).  In the second case, the amount
of host contamination is estimated by minimizing the reduced
$\chi^{2}$.  When the SN is almost free from contamination of the host
galaxy, judged from a visual inspection of the imaging and
spectroscopic data, we do not make any corrections for the host galaxy.

We quantify the degree of confidence that we have correctly classified
a candidate as a SN~Ia using the index described in
\citet{howell2005}.  A confidence index (C.I.), ranging from 5 to 0,
is assigned to each SN candidate that was observed within two rest
frame weeks of maximum light. In addition to the spectra, information
from the lightcurve and the host, if available, are used to set this
index. Candidates with 5, 4, 3, 2, 1, and 0 are {\it certain SN~Ia},
{\it highly probable SN~Ia}, {\it probable SN~Ia}, {\it unknown
  object}, {\it probably not a SN~Ia}, and {\it not a SN~Ia},
respectively.  For those candidates that are classified as SNe~Ia with
C.I.~of 5, 4, and 3, we list the best matching template in 
Table~\ref{tab:obssummary2} and plot the candidate and the best
matching template in the figures of the following section. We
also make this comparison for selected candidates with C.I.~of 2. 
Following \citet{howell2005}, SNe~Ia that are assigned a C.I.~of 4
  or 5 are typed as SN~Ia, and those that are assigned a C.I.~of 3 are
  typed as SN~Ia*, which indicates a higher degree of uncertainty for
  these SNe. No classification is made for candidates that were
observed more than two weeks after maximum light, where the host
galaxy was the primary target.

As a cross check of the classification, all spectra were fitted
with {\it superfit} \citep{howell2005} and independently classified by two
of the authors. In nearly all cases, the classifications
match. For those few cases in which there was disagreement, always small,
we usually adopted the more conservative classification.

The fitting results are summarized in Table~\ref{tab:obssummary2}. 

\section{SN Type Classification Results}\label{sec:sntyping}

In the following sections, we show spectra of 8 candidates from the
Subaru 2001 and Spring 2002 search campaigns, 2 spectra from the Fall 2002
search campaign in the SXDS fields, and 10 spectra from the HST Cluster Supernova Survey. 
The confidence in the spectral classifications varies from 5,
a secure SNe~Ia (SDF1, for example), to 0, not a SN~Ia (SCP06F6, for example).
The spectra of candidates that are clear AGN (C02-007
and SCP06V6, for example) are not shown. Also, the FOCAS spectrum of
the extraordinary transient SCP06F6 was presented in \cite{barbary2009}
and is not re-shown here.  All the candidates observed with FOCAS are
summarised in Tables \ref{tab:obssummary1} and \ref{tab:obssummary2}.

For display purposes only, the spectra shown in the following
figures were smoothed with a Gaussian (FWHM of $\sim30$ pixels or
$\sim40$\AA) with each point weighted with the inverse of its
variance. Although telluric absorption features were removed using
the spectra of bright stars, the correction is not perfect, so these
regions are highlighted in gray. Also highlighted are residuals from
the strong atmospheric emission lines, such as those from Oxygen
($5577$\AA\ and $6300$\AA) and Sodium ($5890$\AA).  Non-stellar
features, such as the [O\emissiontype{II}] emission line, which are
not reproduced by BC03 models, are also marked.  These regions can
bias spectral fitting results, so they are masked out. The
Suprime-Cam and ACS finding charts are $8\arcsec\times8\arcsec$ and
$5\arcsec\times5\arcsec$, respectively.  North is up and East is to the left.

\subsection{Candidates from the Subaru Spring Searches}\label{sec:sninthesubaruspringcampaigns}

All the SN candidates shown in this section were found with Subaru
Suprime-Cam in the springs of 2001 and 2002.

\subsubsection{CL1604\_0-1}\label{sec:cl1604_0-1}

\begin{figure}
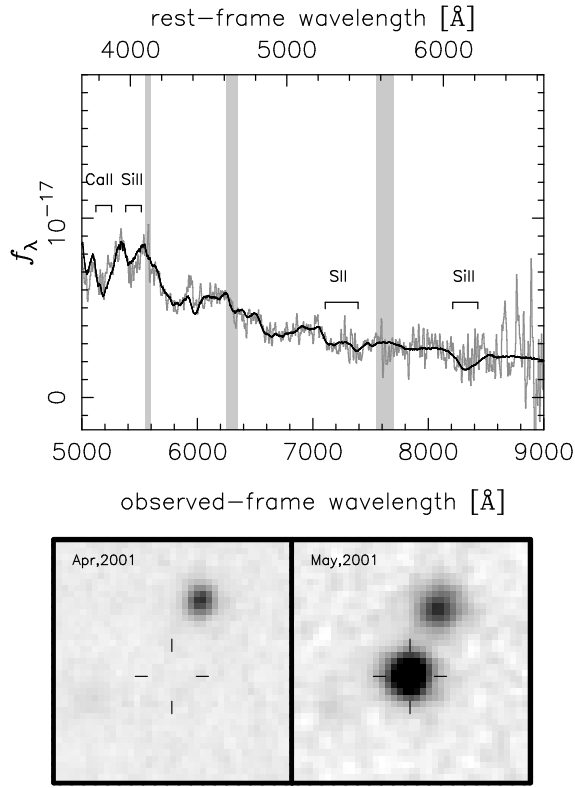

  \begin{center}
\FigureFile(75mm,10mm){./figure1a.eps}
\FigureFile(65mm,10mm){./figure1b.eps}
  \end{center}
\caption{ (Top): The spectrum of CL1604\_0-1 (SN~2001cq) at $z_{\rm{SN}}=0.36$ (gray) 
and the redshifted spectrum of SN~1998aq at $t=-8$ day (black).
The host galaxy redshift $z_{\rm{gal}}$ can not be determined. 
Light gray regions are masked because of atmospheric absorption 
or missubtraction of the sky background. The C.I. is 5. 
(Bottom): Finding charts of CL1604\_0-1. The size is $8\arcsec\times8\arcsec$. 
North is up and east is left. 
\label{fig:cl1604_0-1}}
\end{figure}

CL1604\_0-1 (SN~2001cq) was found in the field surrounding the galaxy
cluster CL1604.  On May 27, 2001, we obtained the spectrum with an
exposure of 900 s.  The host galaxy can not be detected in the images. 
The spectrum (Figure~\ref{fig:cl1604_0-1}) does not show any significant
spectral features from the host galaxy and $z_{\rm{gal}}$ can not be
determined.  Clear Si\emissiontype{II} 4000\AA, S\emissiontype{II}
``w'', and Si\emissiontype{II} 6150\AA\ features are detected,
indicating that this is a secure SN~Ia at $z_{\rm{SN}}=0.36$.  The
C.I. is 5.  The best-fit SN spectrum is a normal SN~Ia, SN~1998aq, at
$t=-8$ days.

\subsubsection{1520.1-2}\label{sec:1520.1-2}

\begin{figure}[htbp!]
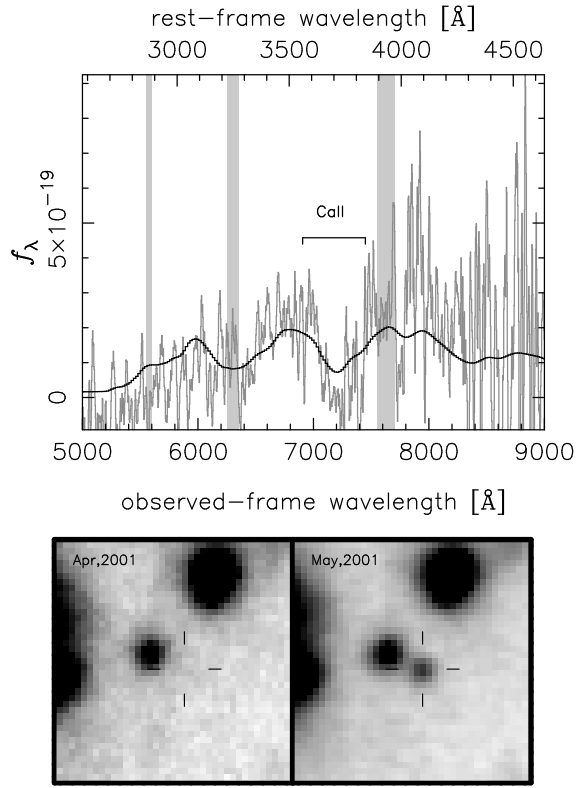

\begin{center}
\FigureFile(75mm,10mm){./figure2a.eps}
\FigureFile(65mm,10mm){./figure2b.eps}
\caption{
(Top):  The spectrum of 1520.1-2 at $z_{\rm{gal}}\sim0.94$ (gray) 
and the Hsiao Ia template at $t=-6$ days (black) redshifted to $z_{\rm{SN}}=0.95$. 
The C.I. is 3. 
(Bottom): Finding charts of 1520.1-2. 
\label{fig:1520.1-2}}
\end{center}
\end{figure}

1520.1-2 (SN~2001cw) was found in the field centered on the galaxy
cluster 1520.1.  The FOCAS spectrum (Figure~\ref{fig:1520.1-2}) was
obtained on May 27, 2001 and the exposure was 4200 seconds.  The SN is
well separated from the probable host, which is located to the
northeast of the SN.  The redshift of this galaxy is
$z_{\rm{gal}}\sim0.94$ as determined from the 4000\AA\ break.  The
spectrum can fitted with both the Nugent Type~Ib/c template at $t=-4$
days redshifted to $z_{\rm{SN}}=0.95$ and the Hsiao Type~Ia template at $t=-6$ days redshifted to
$z_{\rm{SN}}=0.95$.  The light curve is consistent with that of a SN
Ia (\cite{amanullah2009}) and also $t_{\rm{sp}}$ matches $t_{\rm{LC}}$. 
We conclude that it is a probable SN~Ia and set the C.I. to 3.

\subsubsection{SDF1}\label{sec:sdf1}

\begin{figure}[htbp!]
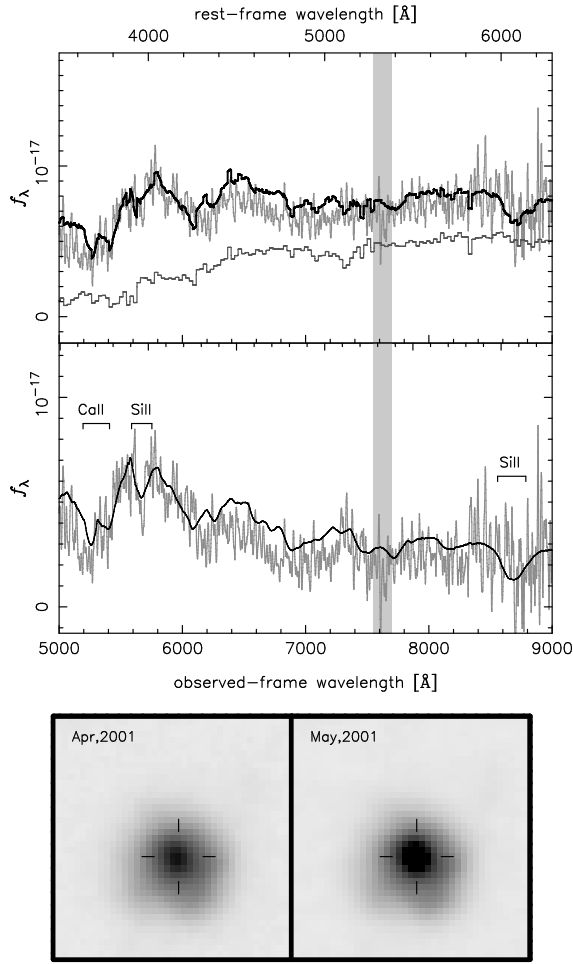

\begin{center}
\FigureFile(75mm,10mm){./figure3a.eps}
\FigureFile(65mm,10mm){./figure3b.eps}
\caption{
(Top): In the upper panel, the spectrum of SDF1 (SN~2001cs) at $z_{\rm{gal}}=0.431$ (gray), 
the galaxy template subtracted spectrum (dark gray), 
and the combined SN template plus galaxy template spectrum (black).
In the lower panel, the galaxy-subtracted spectrum (gray) and 
and the spectrum of the best matching local SN, SN~1994D 
at $t=-1$ days, redshifted to $z_{\rm{SN}}=0.42$ (black). 
The C.I. is 5.
(Bottom): Finding charts of SDF1. 
\label{fig:sdf1}}
\end{center}
\end{figure}

SDF1 (SN~2001cs) was found during the 2001 run of the SDF project
together with other SDF SN candidates (SDF2, SDF4, SDF5, and SDF6).
This SN is only slightly offset from the center of the relatively
bright host. Not surprisingly, the SN spectrum is strongly
contaminated by host galaxy light.  The 600 second spectrum (Figure
\ref{fig:sdf1}) taken on May 26, 2001 indicates that the host galaxy is
at $z_{\rm{gal}}=0.431$ from Ca\emissiontype{II} H and K absorption lines.
After subtracting a model of the host galaxy spectrum and
correcting for extinction (E$(B-V)=0.2$ mag), the distinctive Si
\emissiontype{II} feature of SNe~Ia around 4000\AA\ in the rest frame
can be seen. This is a SN~Ia with the C.I. of 5 at
$z_{\rm{SN}}=0.42$.  The best-fit spectrum is SN~1994D at $t=-1$ days.

\subsubsection{SDF2}\label{sec:sdf2}

\begin{figure}[htbp!]
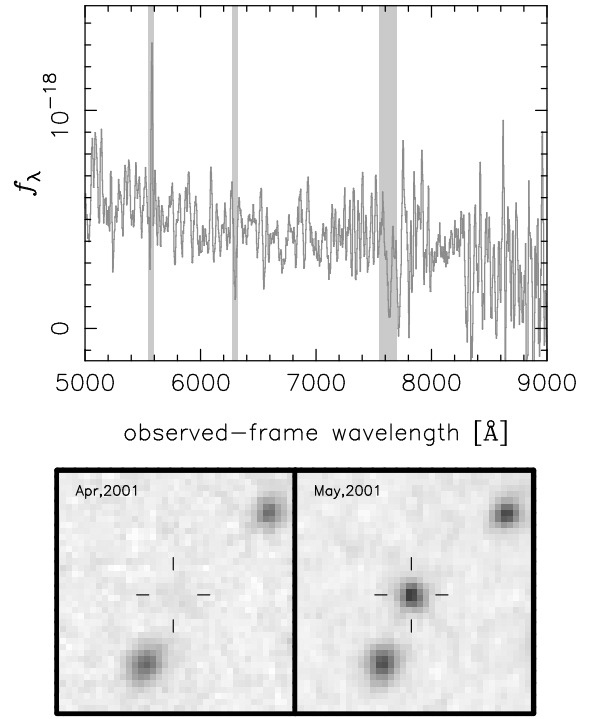

\begin{center}
\FigureFile(75mm,10mm){./figure4a.eps}
\FigureFile(65mm,10mm){./figure4b.eps}
\caption{
(Top): Spectrum of SDF2 (SN~2001ct). 
The C.I. is 1.
 (Bottom): Finding charts of SDF2. 
\label{fig:sdf2}}
\end{center}
\end{figure}

The spectrum of SDF2 (SN~2001ct) is shown in Figure~\ref{fig:sdf2}.
It was taken on May 27, 2001 and has a total integration time of 6300
seconds.  This SN appears on a diffuse galaxy.  The redshift of
the likely host could not be determined and the spectrum of the SN is
featureless and blue. 
We can determine neither
the type nor the redshift of SDF2; however, the blue continuum
suggests that this candidate is probably a Type~II SNe. The C.I. is 1.

\subsubsection{SDF4}\label{sec:sdf4}

\begin{figure}[htbp!]
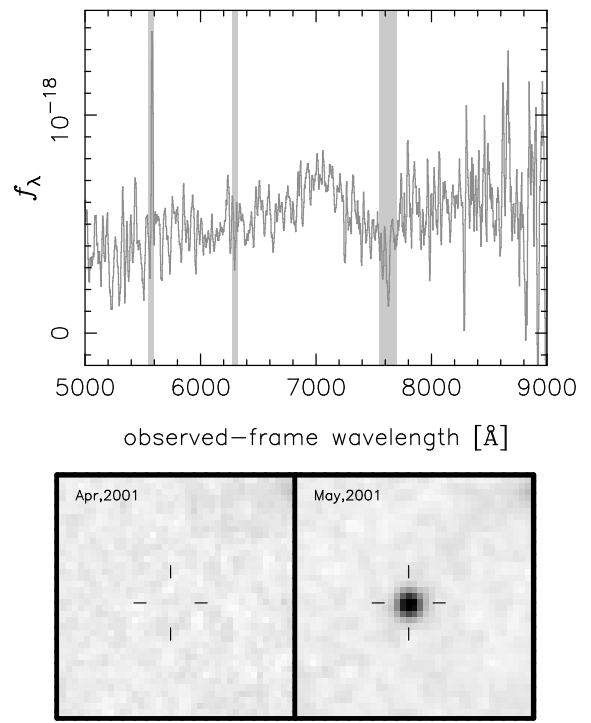

\begin{center}
\FigureFile(75mm,10mm){./figure5a.eps}
\FigureFile(65mm,10mm){./figure5b.eps}
\caption{
(Top): 
Spectrum of SDF4 (SN~2001cr). 
The C.I. is 2.
(Bottom): Finding charts of SDF4. 
\label{fig:sdf4}}
\end{center}
\end{figure}

The spectrum of the unusual transient SDF4 (SN~2001cr) is shown in
Figure~\ref{fig:sdf4}. It was taken on May 27, 2001 and has a total
integration of 2100 seconds.  The Suprime-Cam reference image taken in
April 2002 does not show any significant signal from the either the SN
or the host.  The SN component dominates the spectrum. The redshift of
the host galaxy $z_{\rm{gal}}$ could not be determined.  There are no
SN templates that fit the observed spectrum well.  The best-fit SN
template is the Nugent SN~Ib/c hypernova (HN) template at $t=0$ days redshifted
$z_{\rm{SN}}=0.78$.  But some bumpy structures in the spectrum do not
match well.  A SN~1991T template at $t=-11$ days redshifted to
$z_{\rm{SN}}=0.81$ might fit the spectrum better although the $\chi^2$
values are slightly larger. The C.I. is 2.

\subsubsection{SDF5}\label{sec:sdf5}

The spectrum of SDF5 (SN~2001cv) is shown in Figure~\ref{fig:sdf5}.
It was taken on May 26, 2001 and has a total integration of 5300
seconds.  The SN is separated from the host galaxy and the SN
component dominates the spectrum at the location of the SN.  The
redshift of the host is $z_{\rm{gal}}=1.039$ as measured from 
a single strong emission line, which we identify as
[O\emissiontype{II}]. It is very unlikley that the line is
Ly$\alpha$, as the line is not asymmetric and the continuum across
the line shows no evidence for a discontinuity. It is unlikely
to be H$\alpha$ or [O\emissiontype{III}] as no other lines are seen,
despite the strength of the detected line. 
When the spectral template of a star
forming galaxy is subtracted, the SN spectral fitting indicates that
the Hsiao template SN~Ia at $t=9$ days at $z_{\rm{SN}}=1.02$ is the
best match.  However, the spectral features characteristic of a SN~Ia
are marginal and we do not have a well-sampled light curve.
Therefore, we set the C.I. to 2.

\begin{figure}[htbp!]
\begin{center}
\FigureFile(75mm,10mm){./figure6a.eps}
\FigureFile(65mm,10mm){./figure6b.eps}
\caption{
(Top): The spectrum of SDF5 (SN~2001cv) at $z_{\rm{gal}}=1.039$ (gray) 
and the Hsiao SN~Ia template at $t=9$ days redshifted to $z_{\rm{SN}}=1.02$ (black). 
The C.I. is 2. 
(Bottom): Finding charts of SDF5. 
\label{fig:sdf5}}
\end{center}
\end{figure}

\subsubsection{SDF6}\label{sec:sdf6}

The spectrum of SDF6 (SN~2001cu) is shown in Figure~\ref{fig:sdf6}.
It was taken on May 27, 2001 and has a total integration time of 900
seconds.  The host is similar in brightness to the SN, so the SN
spectrum is contaminated by light from the host.  A redshift of
$z_{\rm{gal}}=0.515$ is measured from the Ca\emissiontype{II} H and K
absorption lines of the host galaxy.  After subtracting a
  passively evolving model for the spectrum of the host galaxy and
  correcting for extinction ($E(B-V)=0.2$ mag) both the Nugent Ib/c
  template at $t=-6$ days redshifted to $z_{\rm{SN}}=0.50$
  and the Hsiao Ia template at $t=-6$ days redshifted to
  $z_{\rm{SN}}=0.50$ are acceptable fits to the spectrum. Since
there a too few light curve points to help with the classification,
the C.I. is set to 2.

\begin{figure}[htbp!]
\begin{center}
\FigureFile(75mm,10mm){./figure7a.eps}
\FigureFile(75mm,10mm){./figure7b.eps}
\FigureFile(65mm,10mm){./figure7c.eps}
\caption{
(Top):  The spectrum of SDF6 (SN~2001cu) at $z_{\rm{gal}}=0.515$ (gray) 
and the Nugent Ib/c template at $t=-6$ days redshifted to $z_{\rm{SN}}=0.50$ (black). 
(Middle): 
Spectrum of SDF6 at $z_{\rm{gal}}=0.515$ (gray) 
and the Hsiao Ia template at $t=-6$ days redshifted to $z_{\rm{SN}}=0.50$ (black). 
The C.I. is 2. 
(Bottom): Finding charts of SDF6. 
\label{fig:sdf6}}
\end{center}
\end{figure}

\subsubsection{S02-032}\label{sec:s02-032}

\begin{figure}[htbp!]
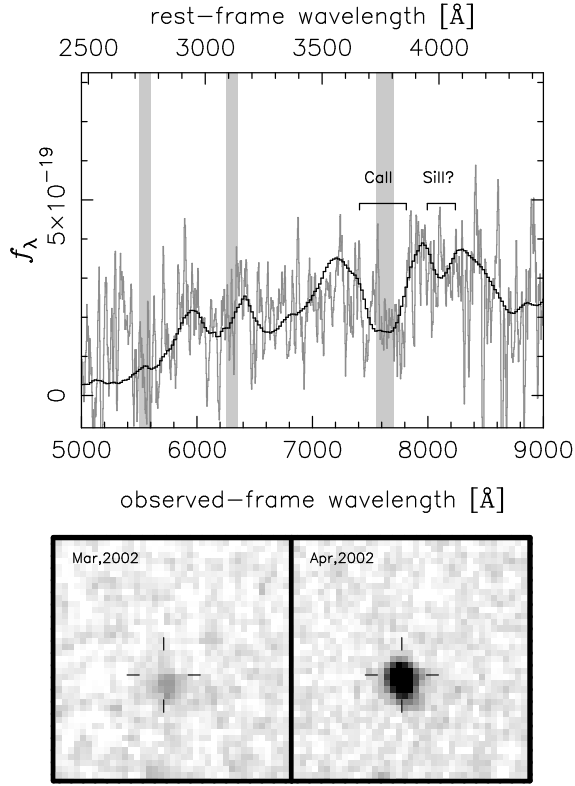

\begin{center}
\FigureFile(75mm,10mm){./figure8a.eps}
\FigureFile(65mm,10mm){./figure8b.eps}
\caption{
(Top): The spectrum of S02-032 (SN~2002ff) at $z_{\rm{SN}}=1.02$ (gray) 
and the Nugent Ia template at $t=0$ days (black) redshifted to $z_{\rm{SN}}=1.02$. 
The C.I. is 4. 
(Bottom): Finding charts of S02-032. 
\label{fig:s02-032}}
\end{center}
\end{figure}

The spectrum of S02-032 (SN~2002ff) is shown in Figure~\ref{fig:s02-032}. 
It was taken on April 17, 2002 and has a total
integration time of 5400 seconds.  This SN appears on a diffuse galaxy
whose redshift could not be measured.  The SN component dominates the
spectrum.  The SN spectral fitting indicates that this is a SN~Ia at
$z_{\rm{SN}}=1.02$.  The best-fit spectrum template is the Nugent normal
Ia template at $t=0$ days.  The light curve is consistent with that of
a SN~Ia and $t_{\rm{sp}}$ matches $t_{\rm{LC}}$. The C.I. is 4.

\subsection{Candidates in the SXDS field}\label{sec:sninthesxds}

All the SN candidates shown in this section were found with Subaru
Suprime-Cam during the SN campaign conducted during the Fall of 2002.

\subsubsection{SuF02-012}\label{sec:suf02-012}

\begin{figure}[htbp!]
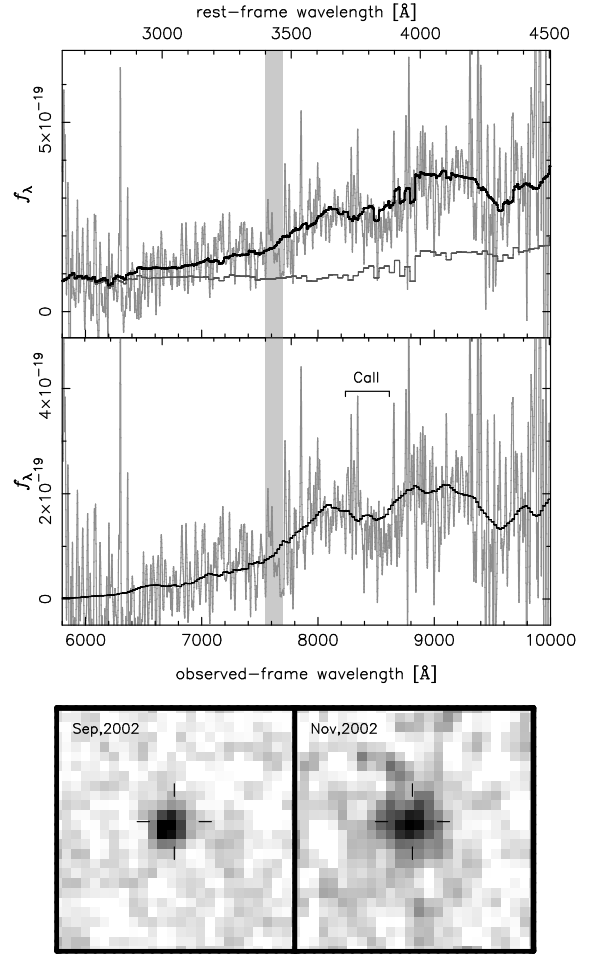

\begin{center}
\FigureFile(75mm,10mm){./figure9a.eps}
\FigureFile(65mm,10mm){./figure9b.eps}
\caption{
(top): 
Spectra of SuF02-012 (SN~2002lc) at $z_{\rm{SN}}=1.22$ (gray) and 
SN~1991T at $t=12$ days (black) redshifted to $z_{\rm{SN}}=1.22$. 
The C.I. is 2. 
(bottom): Finding charts of SuF02-012. 
\label{fig:suf02-012}}
\end{center}
\end{figure}

The spectrum of SuF02-012 (SN~2002lc) is shown in Figure
\ref{fig:suf02-012}. It was observed on November 13, 2002 and has a
total integration time of 7200 seconds.  It is hosted by a
  diffuse galaxy, whose redshift could not be determined.  The
  spectral fitting indicates that it might be a SN~Ia at
  $z_{\rm{SN}}=1.22$.  However, the phase of the best-fit template, SN
  1991T at $t=12$ days, disagrees with the light curve phase. SNe~Ia at
  earlier phases and higher redshifts are also a possibility.  Given the
  uncertainty, we have set the C.I.~to 2. This target was a high
priority object and spectroscopically observed with FORS2 on the ESO
VLT, ESI on the Keck, and the ACS grism on the HST.  All the spectra
suggest that this is a high redshift SN.  A future analysis of the
combined data set may lead to this candidate being reclassified.

\subsubsection{SuF02-061}\label{sec:suf02-061}

\begin{figure}[htbp!]
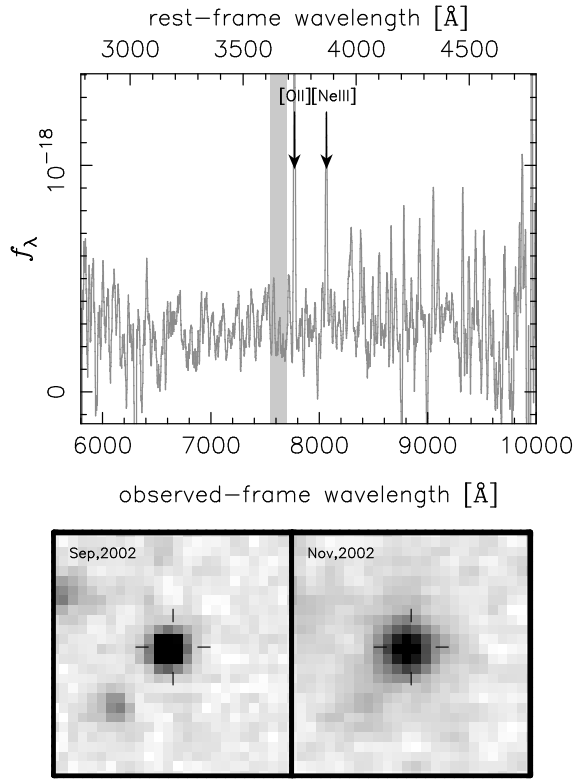

\begin{center}
\FigureFile(75mm,10mm){./figure10a.eps}
\FigureFile(65mm,10mm){./figure10b.eps}
\caption{
(top): 
Spectra of SuF02-061 at $z_{\rm{gal}}=1.085$. 
The C.I. is 1. 
(bottom): Finding charts of SuF02-061. 
\label{fig:suf02-061}}
\end{center}
\end{figure}

The spectrum of SuF02-061 is shown in Figure~\ref{fig:suf02-061}. It
was taken on November 13, 2002 with a total exposure time of 5400
seconds.  This SN appears on a diffuse galaxy that has a redshift of
$z_{\rm{gal}}=1.085$, as measured from several emission lines,
including [O\emissiontype{II}], [Ne\emissiontype{III}], and some
Balmer lines. 
The spectrum is too noisy or the host light dominates too much 
to classify the SN type. 
This galaxy is also detected in X-rays, suggesting that the variability may 
originate from AGN activity. The C.I. is 2.

\subsection{Candidates from the HST Cluster Supernova Survey}\label{sec:sninthehstclustersn}

All the SN candidates shown in this section were found in the HST ACS
survey for SNe~Ia in distant galaxy clusters (HST Cluster Supernova
Survey).  On average, the SN candidates in this search are more
distant than the ones found in earlier searches. Furthermore, since
the survey aimed to find SNe~Ia in 
early-type galaxies\footnote{If a galaxy has no detectable 
star formation and has the morphology of an early type galaxy, 
then we use the term
early-type to describe the galaxy. It is a definition that is 
commonly used throughout the literature. To our knowledge, there has 
never been a Type II Supernova in a galaxy where there was no 
detectable signs of star formation. Of course, some
star formation may be beyond detection in our optical spectra
because it is either heavily obscured or because the star formation
rate is very low.}, the relative
brightness of such SN with respect to the host is usually
small. Consequently, typing of these candidates generally needs to
rely on other methods, such as the properties of the host. The
principle aim of the spectroscopy is to get the redshift of the host.
Note that the finding charts have been made from images taken
with ACS. They have greater spatial resolution than those made from
images taken with Suprime-Cam, which are smoothed by the seeing.

\subsubsection{SCP06B3}\label{sec:b-003}

\begin{figure}[htbp!]
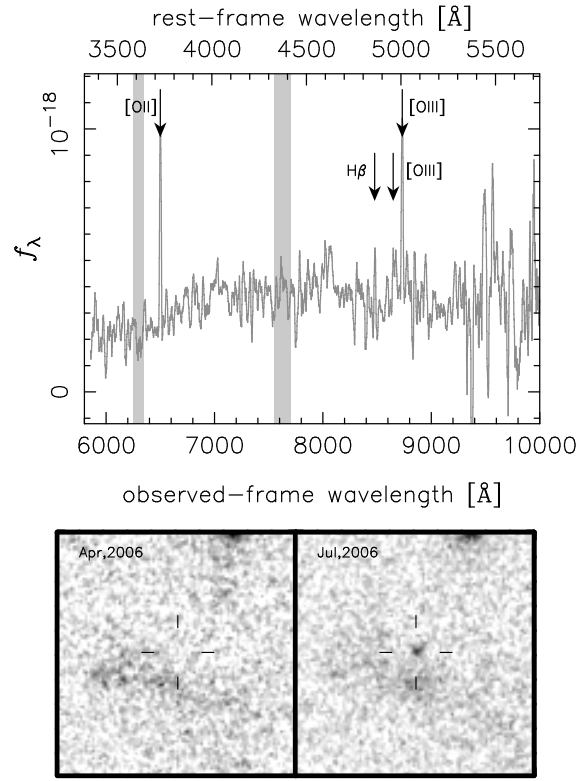

\begin{center}
\FigureFile(75mm,10mm){./figure11a.eps}
\FigureFile(65mm,10mm){./figure11b.eps}
\caption{
(Top): 
Spectrum of SCP06B3 at $z_{\rm{gal}}=0.744$ (gray). 
The C.I. is 2. 
(Bottom): Finding charts of SCP06B3. 
The size is $5\arcsec\times5\arcsec$. North is up and east is left. 
\label{fig:b-003}}
\end{center}
\end{figure}

The spectrum of SCP06B3 is shown in Figure~\ref{fig:b-003}. It was
taken on August 22, 2006 and has a total integration time of 7200
seconds.  A second candidate, SCP06B4 (\S\ref{sec:b-004}), was
observed in the same slitmask.  SCP06B3 appears near to a very diffuse
galaxy that has a redshift of $z_{\rm{gal}}=0.744$ from
[O\emissiontype{II}], H$\beta$, and [O\emissiontype{III}] emission lines. 
The spectrum is dominated by host galaxy light and a
convincing type could not be obtained. The C.I.~is set to 2.

\subsubsection{SCP06B4}\label{sec:b-004}

\begin{figure}[htbp!]
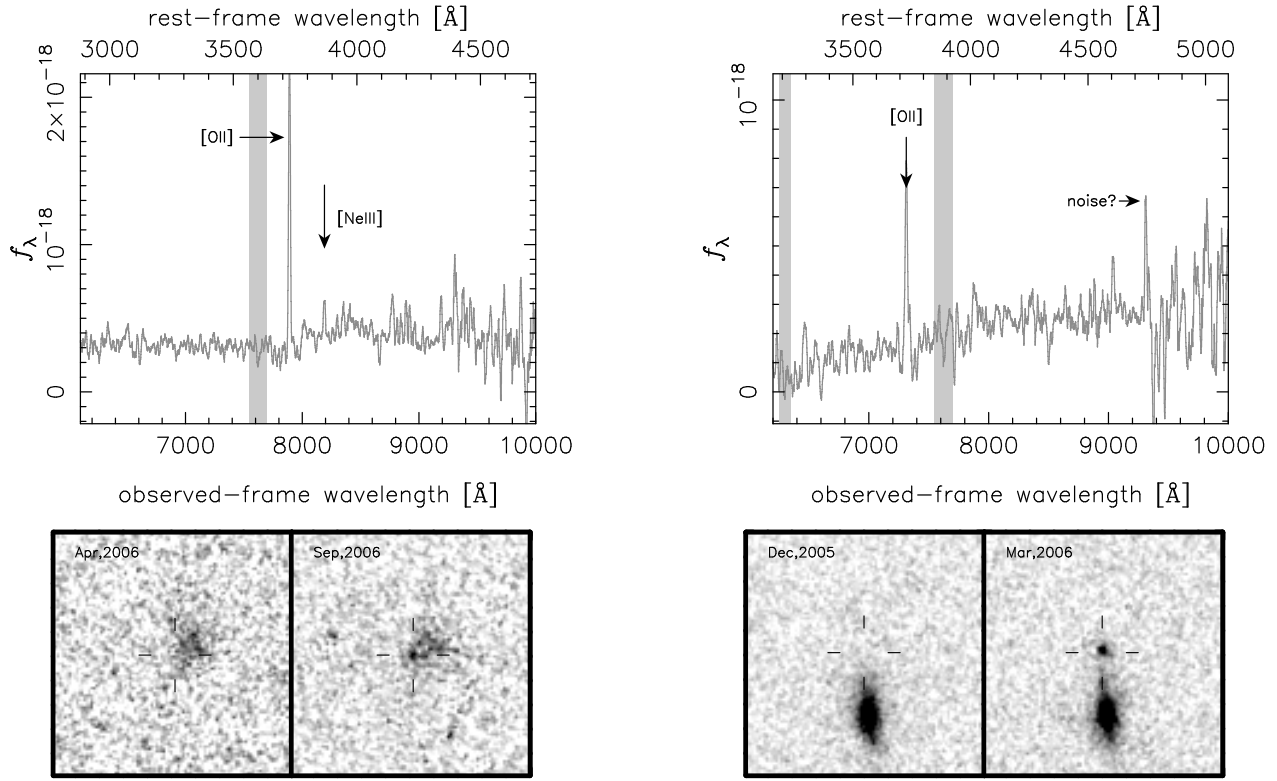

\begin{center}
\FigureFile(75mm,10mm){./figure12a.eps}
\FigureFile(65mm,10mm){./figure12b.eps}
\caption{
(Top): 
Spectrum of SCP06B4 at $z_{\rm{gal}}=1.117$ (gray). 
The C.I. is 2. 
(Bottom): Finding charts of SCP06B4. 
\label{fig:b-004}}
\end{center}
\end{figure}

The spectrum of SCP06B4 is shown in Figure~\ref{fig:b-004}. It was
taken on August 22, 2006 and has a a total integration time of 7200
seconds.  SCP06B4 appears near to a diffuse galaxy that has a redshift
of $z_{\rm{gal}}=1.117$ as measured from [O\emissiontype{II}] and [Ne\emissiontype{III}]
emission lines. The spectrum of SCP06B4 appears to be dominated by light from
the host. A convincing type for the transient could not be derived, so the
C.I.~is set to 2.

\subsubsection{SCP06F6} 
\label{sec:f-006}

The extraordinary transient SCP06F6 was found on February 21, 2006.
It shows a large (from below the detection limit in the ACS $i_{775}$
and $z_{850}$ bands) and slow ($\sim5$ months in the observer frame)
increase in flux \citep{dawson2006}.  For this interesting and unusual
object, we took FOCAS spectra three times on April 22 2006 (6000 sec),
June 28 2006 (4800 sec), and August 22 2006 (2400 sec).  However, the
signal-to-noise ratio of the last two spectra are too low to see any spectral
features.  SCP06F6 seems to be hostless.  The properties of this
object are discussed in detail in \cite{barbary2009}. It is quite unlike
a SN~Ia, so the C.I. is set to 0.

\subsubsection{SCP06G3}\label{sec:g-003}

The spectrum of SCP06G3 is shown in Figure~\ref{fig:g-003}.  It was
taken on April 23 and 24, 2006 and has a total exposure time of 22,800
seconds.  A second candidate, SCP06G4 (\S\ref{sec:g-004}), was
observed in the same slitmask.  SCP06G3 appears to be associated with
a galaxy that can be seen to the south of SCP06G3 (see
Figure~\ref{fig:g-003}).  A single strong emission line, which we
identify as [O\emissiontype{II}] from the host galaxy, leads to a
redshift of $z_{\rm{gal}}=0.962$. Weak Balmer lines and a weak
4000\AA\ break are also detected. The position angle of this
slitmask was optimized for SCP06G4, which meant that the slit could
not be set along the line joining SGP06G4 and the center of the host
galaxy.  Also, the spectrum was taken 24 days after maximum light, so
it consists mostly of light from the host galaxy. For these reasons,
this candidate could not be typed and a C.I.~was not assigned.

\begin{figure}[htbp!]
\begin{center}
\FigureFile(75mm,10mm){./figure13a.eps}
\FigureFile(65mm,10mm){./figure13b.eps}
\caption{
(top): 
Spectrum of SCP06G3 at $z_{\rm{gal}}=0.962$ (gray). 
The sharp feature at 9300\AA\ is a noise artifact from
the processing of the data and is not real. 
The C.I. is 2. 
(bottom): Finding charts of SCP06G3. 
\label{fig:g-003}}
\end{center}
\end{figure}

\subsubsection{SCP06G4}\label{sec:g-004}

\begin{figure}[htbp!]
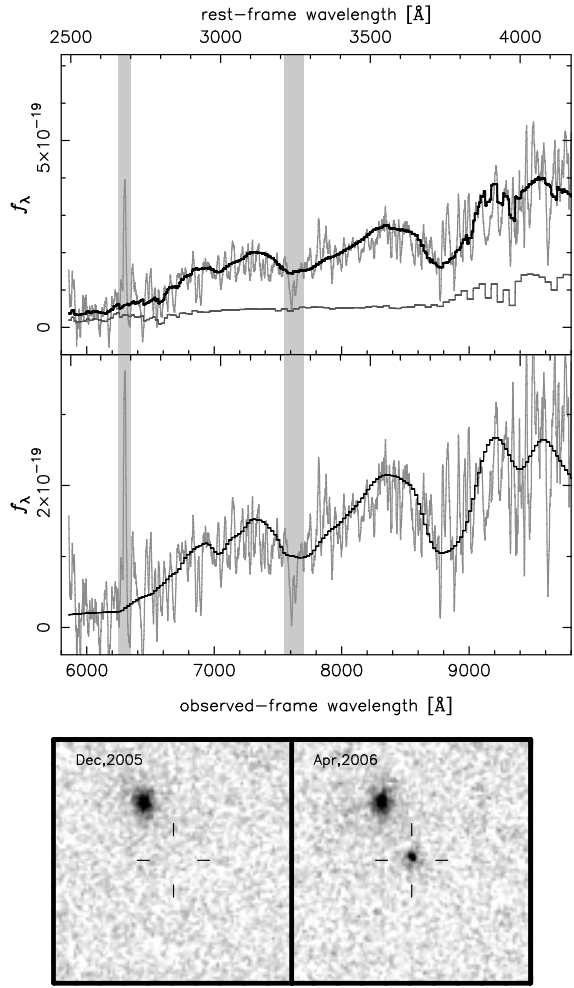

\begin{center}
\FigureFile(75mm,10mm){./figure14a.eps}
\FigureFile(65mm,10mm){./figure14b.eps}
\caption{
(top): 
The spectrum of SCP06G4 at $z_{\rm{gal,fit}}=1.350$ (gray) and 
the Hsiao Ia template at $t=-1$ days redshifted to $z_{\rm{SN}}=1.35$ (black). 
The C.I. is 3. 
(bottom): Finding charts of SCP06G4. 
\label{fig:g-004}}
\end{center}
\end{figure}

The spectrum of SCP06G4 is shown in Figure~\ref{fig:g-004}. It was
taken on April 22 and 23, 2006 and the total integration was 22800
seconds.  SCP06G4 is well separated from the putative host galaxy which lies
to the northeast of SCP06G4. The redshift of the host is
$z_{\rm{gal}}\sim1.35$ from Ca\emissiontype{II} H and K absorption lines
and the 4000\AA\ break. A spectral fit to the host galaxy results in a
similar redshift, $z_{\rm{gal}}=1.350$ (see
\S\ref{sec:focashostspectraforhstclustersn}). 

Although SCP06G4 is well separated from the host galaxy in the ACS
images, the SN spectrum was strongly contaminated by light from the
host galaxy. 
Thus, we subtract a spectrum that was extracted from a
location that is at the same distance from the galaxy center to the SN
but on the other side of the galaxy. The host subtracted spectrum of
SCP06G4 is very noisy and can be fitted with a SN~Ib/c template at
$t=-6$ days redshifted to $z_{\rm{SN}}=1.36$ or the Hsiao Ia template at
$t=-1$ days redshifted to $z_{\rm{SN}}=1.35$. 
Host galaxy subtraction using the BC03 templates also gives 
the best fitting result of $t=-1$ days redshifted to $z_{\rm{SN}}=1.35$ 
(the upper panel of Figure \ref{fig:g-004}). 
The light curve is
consistent with that of a SN~Ia and $t_{\rm{sp}}$ matches
$t_{\rm{LC}}$. The C.I. is 3. The classification is supported by
the characteristics of the host galaxy. The morphology of host is
consistent with it being an elliptical galaxy (Meyers el al. in
preparation) and no [O\emissiontype{II}] emission is detected 
in the spectrum of the host.

SCP06G4 is the most distant spectroscopically confirmed SN~Ia in
our sample and the most distant spectroscopically confirmed SN~Ia
using a ground based telescope. The confidence in the classification
is similar to that that can be obtained for other similarly distant
SNe that have ACS grism spectra (\cite{riess2007}). Clearly, in some
cases, one can do as well from the ground as from space, if
atmospheric conditions are optimal, the host either faint enough or
far enough away from the SN, integrations are long, and some care in
adopting the optimal observing and data reduction strategies are 
employed.

\subsubsection{SCP06H3}\label{sec:h-003}

\begin{figure}[htbp!]
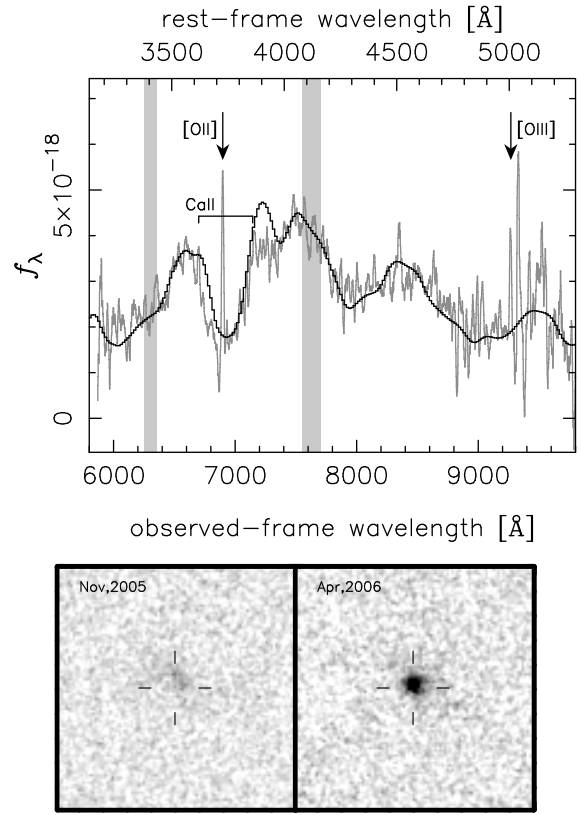

\begin{center}
\FigureFile(75mm,10mm){./figure15a.eps}
\FigureFile(65mm,10mm){./figure15b.eps}
\caption{
(top): 
The spectrum of SCP06H3 at $z_{\rm{gal}}=0.851$ (gray) and 
the Hsiao SN~Ia template at $t=2$ days redshifted to $z_{\rm{SN}}=0.84$ (black). 
The C.I. is 4. 
(bottom): Finding charts of SCP06H3. 
\label{fig:h-003}}
\end{center}
\end{figure}

The spectrum of SCP06H3 is shown in Figure~\ref{fig:h-003}.  It was
taken on April 22, 2006 and the total integration time was 6000
seconds.  SCP06H3 appears to be associated with a very diffuse galaxy.
The redshift of the galaxy is $z_{\rm{gal}}=0.851$ as measured from
[O\emissiontype{II}] and [O\emissiontype{III}].  The spectrum of
SCP06H3 is well fitted to the Hsiao SN~Ia template at $t=2$ days
redshifted to $z_{\rm{SN}}=0.84$.   The light curve is consistent with
that of a SN~Ia and $t_{\rm{sp}}$ matches $t_{\rm{LC}}$, so the C.I. is set to 4.

\subsubsection{SCP06N32}\label{sec:n-032}

\begin{figure}[htbp!]
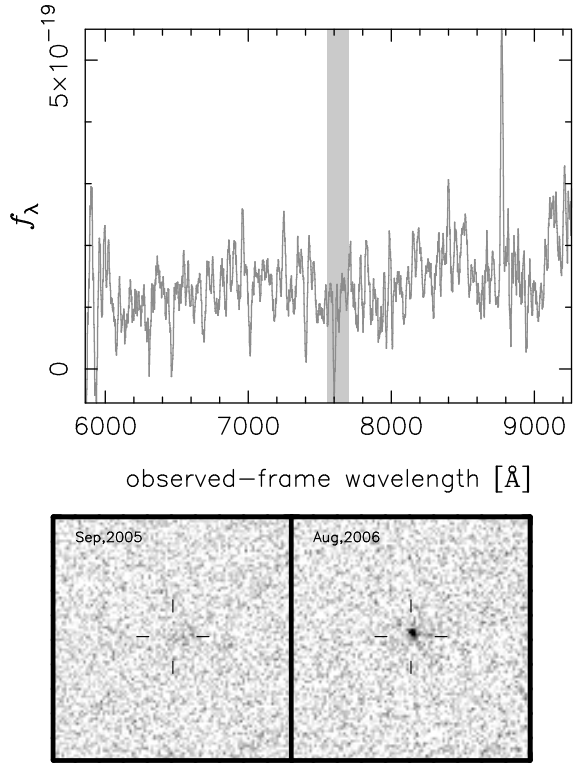

\begin{center}
\FigureFile(75mm,10mm){./figure16a.eps}
\FigureFile(65mm,10mm){./figure16b.eps}
\caption{
(top): 
Spectrum of SCP06N32.
The C.I. is 2. 
(bottom): Finding charts of SCP06N32. 
\label{fig:n-032}}
\end{center}
\end{figure}

The spectrum of SCP06N32 is shown in Figure~\ref{fig:n-032}.  It was
taken on August 22, 2006 and the total integration time was 7200
seconds.  SCP06N32 appears to be associated with a very diffuse and
faint galaxy.  We did not detect any spectral features from the
galaxy.  Although broad features are visible, the spectrum is
noisy and the fitting results are inconclusive.
The C.I. is set to 2.

\subsubsection{SCP05P1}\label{sec:p-001}

\begin{figure}[htbp!]
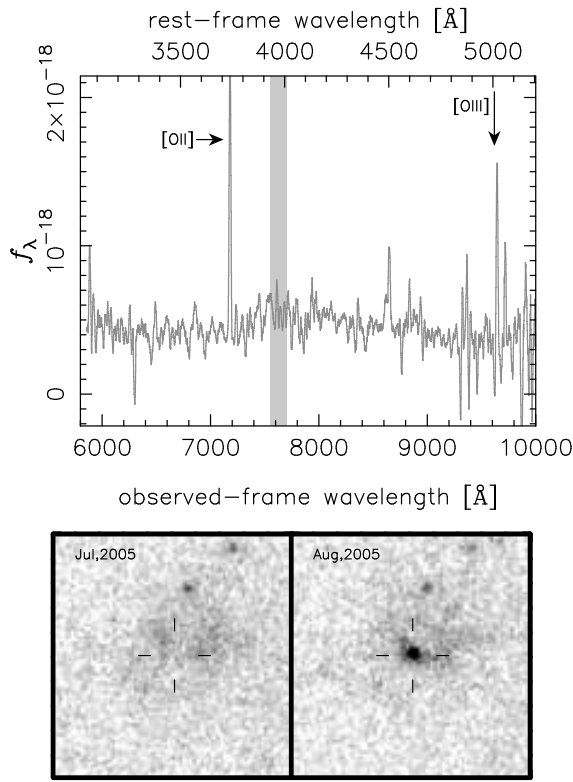

\begin{center}
\FigureFile(75mm,10mm){./figure17a.eps}
\FigureFile(65mm,10mm){./figure17b.eps}
\caption{
(top): 
Spectrum of SCP05P1 at $z_{\rm{gal}}=0.926$ (gray). 
The C.I. is 2. 
(bottom): Finding charts of SCP05P1. 
\label{fig:p-001}}
\end{center}
\end{figure}

The spectrum of SCP05P1 is shown in Figure~\ref{fig:p-001}. It was
taken on September 25, 2005 and the exposure time was 5400 seconds.
SCP05P1 appears near to the center of a very diffuse galaxy, which has
a redshift of $z_{\rm{gal}}=0.926$ from a [O\emissiontype{II}]
emission line.  Since the spectrum was taken $t\sim15$ days after the
light curve peak, we restrict the spectral fits to late SN epochs.
The best fit spectral template is SN~1998S, a Type~IIn SN, at $t=27$
days redshifted to $z_{\rm{SN}}=0.91$, but the fit is poor. Allowing
for some host galaxy light in the spectrum did not lead to a more
conclusive fit. Since the spectrum was taken more than two weeks
after maximum light, the confidence index is left unassigned.

\subsubsection{SCP05P9}\label{sec:p-009}

\begin{figure}[htbp!]
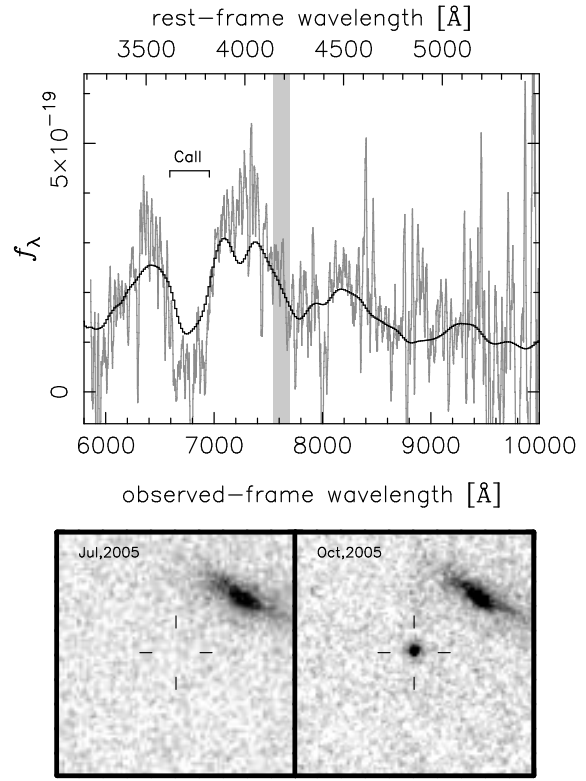

\begin{center}
\FigureFile(75mm,10mm){./figure18a.eps}
\FigureFile(65mm,10mm){./figure18b.eps}
\caption{
(top):  The spectrum of SCP05P9 at $z_{\rm{gal}}=0.821$ (gray) and 
the Hsiao Ia template at $t=-2$ days redshifted to $z_{\rm{SN}}=0.81$ (black). 
We note that we determine the redshift from the spectrum of the host galaxy center 
and the [O\emissiontype{II}] emission line can not be seen in the SN spectrum figure.
The C.I. is 3. 
(bottom): Finding charts of SCP05P9. 
\label{fig:p-009}}
\end{center}
\end{figure}

The spectrum of SCP05P9 is shown in Figure~\ref{fig:p-009}.  It was
taken on October 25, 2005 and the exposure time was 12600 seconds. 
SCP05P9 is well separated from a galaxy that has redshift of
$z_{\rm{gal}}=0.821$ as measured from a [O\emissiontype{II}] emission
line.  The host galaxy is located to the northwest of SCP05P9 
in the finding chart of Figure~\ref{fig:p-009}.
The separation of the SN component from the host galaxy center
is about 1 arcsec and, as in SCP06G4, we subtracted the host galaxy
component from the spectrum at the SN position (\S\ref{sec:g-004}).
The best fit template over the observed wavelength range is the Hsiao Ia
template at $t=-2$ days at $z_{\rm{SN}}=0.81$.  The light curve is
consistent with that of a SN~Ia and $t_{\rm{LC}}$ matches
$t_{\rm{SN}}$.  The C.I. is 3.

\section{Host Galaxy Spectral Fitting for SN Candidates in HST Cluster Supernova Survey}
\label{sec:focashostspectraforhstclustersn}

As an alternative to spectroscopy, classification of the SN
from other observables such as the lightcurve or the properties of
the host can, and sometimes must, be employed.
Identifying the host of a SN as an early type galaxy is a
reasonable indication that the SN is a SN~Ia, as SNe of
other types are unlikely to occur in such galaxies. We fitted BC03
models to the hosts of five SN candidates in HST Cluster Supernova
Survey to constrain the spectral properties of the hosts. The
parameters of the fit were metallicity, star formation history (single
burst or exponentially declining), star formation time scale
$\tau_{\rm{sf}}$, stellar age $t_{\rm{age}}$, and extinction.  In the
fitting, we mask the region covering the atmospheric A-band around
7600\AA\ and any possible emission lines since the BC03 models do not
reproduce non-stellar spectral features.  The spectral fitting also
allowed a more accurate estimate of the redshifts of those host
galaxies for which the Ca II H and K absorption lines were detected
with low signal-to-noise ratio.

The FOCAS spectra of the hosts of SCP05D0, SCP05D6, SCP06G4, SCP06K0,
and SCP06K18 and the best fitting galaxy spectra are shown in Figures
\ref{fig:o-000host}-\ref{fig:k-018host}.  We note that SCP06G4 is
spectroscopically identified as a SN~Ia* (\S\ref{sec:g-004}).
Observations for these host galaxies are summarized in Tables
\ref{tab:obssummary1} and \ref{tab:obssummary2}.  In general, the
contribution from the SN in these spectra can be ignored. SCP05D0,
SCP05D6, SCP06K0, SCP06K18 were all re-observed when the contribution
from the SN was negligible.  For SCP06G4, a spectrum located at the
center of the host was extracted. While some SN light may have leaked
into the slit for these two objects, the contribution was estimated to
be negligible for SCP06G4.

All of the observed galaxy spectra are well fitted with the models
from BC03.  Using the constraint that the age of the universe at the
redshifts the galaxies were observed (for example, 6.0 Gyr at
$z\sim1.0$ and 4.7 Gyr at $z\sim1.4$) should be larger than the
respective ages of these galaxies, we obtain ranges (among the 40 best
fitted templates) for the stellar age $t_{\rm{age}}$ and the star
formation time scale $\tau_{\rm{sf}}$ for each galaxy.  The
uncertainties are large because the observed spectra are noisy. The 40
best fitted templates provide roughly the same $\chi^2$ values.  The
stellar ages $t_{\rm{age}}$ are larger than the timescales of star
formation $\tau_{\rm{sf}}$ for all galaxies.  Among the well-fitted
templates, templates with older ages generally have smaller extinction
than those of younger ages.  From our spectral fitting, we cannot
exclude the possibility of significant extinction, but all the spectra
are consistent with early-type galaxies with little extinction. 
The results above are summarized in Table~\ref{tab:fittingresult_host}.

\begin{table*}
\begin{center}
\scriptsize
\caption{Spectral fitting results for SN hosts. The SN names are
  denoted in column 1. Columns 2 and 3 contain the redshifts measured
  from individual spectral features, some of which are not very
  precise due to the noisy nature of the spectra, and the redshifts
  derived from the spectral fitting, respectively. Columns 4 and 5
  contain the fitted ranges of ages and star formation timescales. 
  Both are in Gyr. 
  In columns 6 and 7, the qualitative strengths of the Ca\emissiontype{II} H and K absorption lines 
  and the 4000\AA\ break are shown, respectively.
}
\label{tab:fittingresult_host} 
\begin{tabular}{lrccccc}\hline
\multicolumn{1}{c}{SN} & 
$z_{\rm{gal}}$ &
$z_{\rm{gal,fit}}$ &
$t_{\rm{age}}$ [Gyr] & 
$\tau_{\rm{sf}}$ [Gyr] & 
Ca\emissiontype{II} H and K & 
4000\AA\ break
\\\hline
SCP05D0 & $1.015$    & $1.015$ & 1.0--3.0 & 0.0--0.5 & clear & clear\\
SCP05D6 & $1.315$    & $1.316$ & 1.5--3.0 & 0.0--0.1 & clear & clear\\
SCP06G4 & $\sim1.35$ & $1.350$ & 1.0--1.5 & 0.0--0.3 & exist & clear\\
SCP06K0 & $\sim1.41$ & $1.416$ & 1.5--3.0 & 0.0--0.3 & exist & clear\\
SCP06K18& $\sim1.41$ & $1.411$ & 1.5--3.0 & 0.0--0.5 & exist & clear
\\\hline
\multicolumn{7}{@{}l@{}}{\hbox to 0pt{\parbox{180mm}{\footnotesize
}\hss}}
\end{tabular}
\end{center}
\end{table*}

\begin{figure}[htbp!]
\begin{center}
\FigureFile(75mm,10mm){./figure19.eps}
\caption{
Spectrum of the host of SCP05D0 (gray) and
the best fitting galaxy template at $z_{\rm{gal,fit}}=1.015$ (black). 
\label{fig:o-000host}}
\end{center}
\end{figure}
\begin{figure}[htbp!]
\begin{center}
\FigureFile(75mm,10mm){./figure20.eps}
\caption{
Spectrum of the host of SCP05D6 (gray) and
the best fitting galaxy template at $z_{\rm{gal,fit}}=1.316$ (black). 
\label{fig:o-006host}}
\end{center}
\end{figure}
\begin{figure}[htbp!]
\begin{center}
\FigureFile(75mm,10mm){./figure21.eps}
\caption{
Spectrum of the host of SCP06G4 (gray) and 
the best fitting galaxy template at $z_{\rm{gal,fit}}=1.350$ (black). 
\label{fig:g-004host}}
\end{center}
\end{figure}
\begin{figure}[htbp!]
\begin{center}
\FigureFile(75mm,10mm){./figure22.eps}
\caption{
Spectrum of the host of SCP06K0 (gray) and 
the best fitting galaxy template at $z_{\rm{gal,fit}}=1.416$ (black). 
\label{fig:k-000host}}
\end{center}
\end{figure}
\begin{figure}[htbp!]
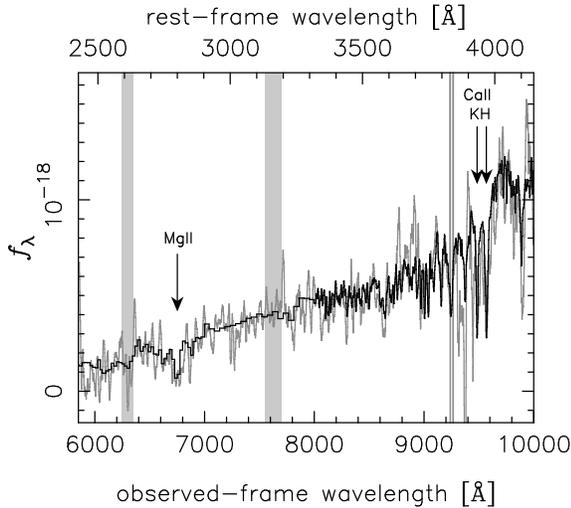

\begin{center}
\FigureFile(75mm,10mm){./figure23.eps}
\caption{
Spectrum of the host of SCP06K18 (gray) and 
the best fitting galaxy template at $z_{\rm{gal,fit}}=1.411$ (black). 
\label{fig:k-018host}}
\end{center}
\end{figure}

\section{Classifying High Redshift Candidates}\label{sec:typeideff}

The chance that any given high redshift candidate is positively
identified as a SN of some type depends on many factors. A
non-exhaustive list includes factors like the redshift of the candidate,
the phase at which it was observed, the contrast with respect to the
host, the wavelength coverage and signal-to-noise ratio of the
spectrum, and, if photometry is used as part of the classification,
the quality and amount of photometry.  The lack of a large sample of
UV spectra for SNe of all types, especially for CC SNe, is also a
factor. 

In Figures 24 and 25 we compare magnitudes, redshifts and
signal-to-noise ratios of candidates that were successfully identified
with those that were not. Only candidates that were observed within 
about two rest frame weeks of maximum light are included in these plots. 
The signal-to-noise ratios include both SN and host galaxy components
except for SCP06G4 and SCP05P9, for which we subtract the spectrum of
the host.  Although exposure times and weather conditions differ from
one candidate to the next, one generally finds, not surprisingly, that
it is easier to identify the SN type when they are at lower redshift,
when they are brighter and when the signal-to-noise of the spectrum is
higher. 

\begin{figure}
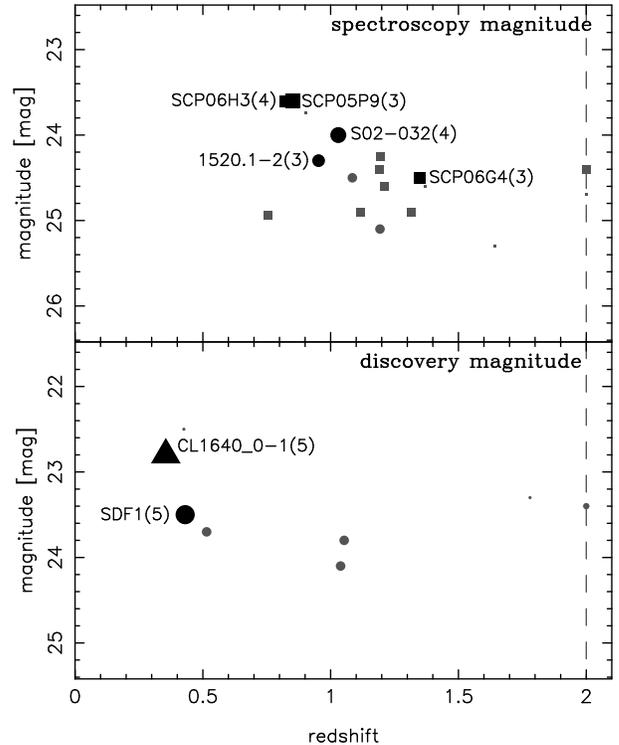

 \begin{center}
\FigureFile(80mm,0mm){./figure24.eps}
 \end{center}
\caption{
The distribution of SN candidate magnitudes as a function of redshift.
If a well-enough sampled lightcurve is available, the magnitude at the
time the spectrum was taken is used and is shown in the upper panel.
Otherwise the magnitude at the time of discovery is used and is shown in
the lower panel. The symbol size is proportional to the degree of confidence that
the candidate is a SN~Ia. 
We plot candidates with C.I.$\leq2$ in gray.
Triangle, circles and boxes represent $R_c$, $i'$, and
$z'-$ magnitudes, respectively.
Symbols on the vertical gray dashed line at $z=2$ are objects whose redshifts
can not be determined. 
\label{fig:sndist00}}
\end{figure}

\begin{figure}
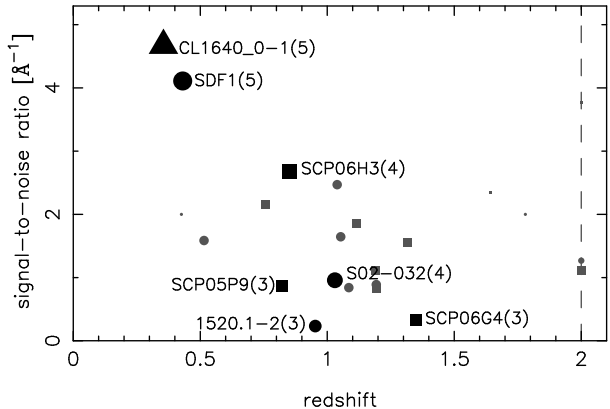

 \begin{center}
\FigureFile(80mm,0mm){./figure25.eps}
 \end{center}\
\caption{
The distribution of signal-to-noise ratios of the spectra as a function of redshift. 
Symbols are the same as those in Figure~\ref{fig:sndist00}. 
For SCP05P9 and SCP06G4, we plot signal-to-noise ratios of host-subtracted SN component spectra. 
Symbols on the vertical gray dashed line at $z=2$ are objects whose redshifts
can not be determined. 
For reference, a 2-hour, low resolution ($R\sim500$) spectrum with FOCAS of a
$z'\sim25$ mag point source has $S/N=1$ [\AA$^{-1}$]. 
The signal-to-noise ratios are not scaled to a common exposure time. 
Doing so tends to stretch the plot vertically. 
\label{fig:sndist01}}
\end{figure}

\begin{figure}
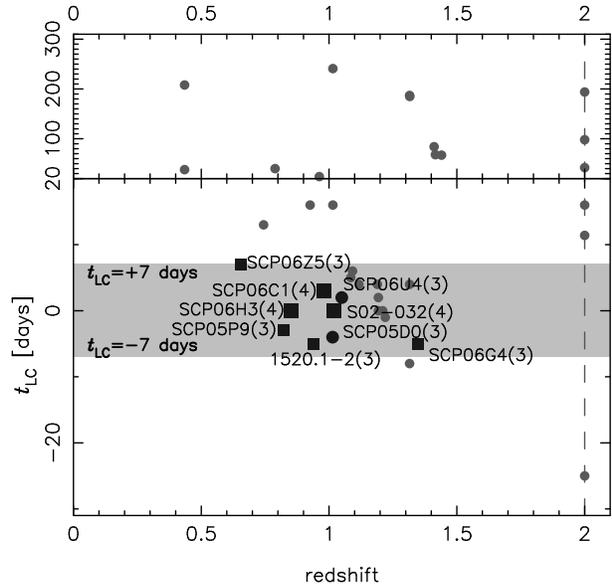

 \begin{center}
\FigureFile(80mm,0mm){./figure26.eps}
 \end{center}
\caption{
The distribution of the epoch the spectra were taken as a function of redshift. 
We only consider objects with well-sampled light curves that have well a determined
date of peak brightness. 
Circles indicate candidates that we could only identify as a SN~Ia when a 
spectrum of the host was available. The gray region is $-7\leq t_{\rm{LC}}\leq7$ days. 
We note that circles at $z=2$ are objects without spectroscopic 
redshift determination and we plot observed-frame epochs for these objects. The object
at $t_{\rm{LC}}\sim-25$ days is the extraordinary transient SCP06F6.
\label{fig:sndistepoch01}}
\end{figure}

Also critical is the phase of the SN when the spectrum is taken.  In
Figure~\ref{fig:sndistepoch01}, $t_{\rm{LC}}$ (the light curve epoch
at the time the spectrum was taken) is plotted as a function of redshift.
Only objects with well-sampled light curves are shown, so objects like
CL1604\_0-1 and SDF1 are excluded.  Black circles and boxes are SNe
that have a C.I. that is greater than 2. 
Considering only the spectra within the shaded region, 5 out of 
11 candidates observed with FOCAS are spectroscopically identified 
as SN~Ia or SN~Ia*. Of
these 5, 4 are below $z=1.1$.  To increase the number of candidates
in this comparison, we add 8 SN candidates from the HST Cluster
Supernova Survey \citep{dawson2009} that were observed with FORS2:
SCP06A4, SCP06C0, SCP06C1, SCP05D0, SCP05D6, SCP06R12, SCP06U4 and
SCP06Z5. Considering only the spectra within the shaded region and
combining the results from FOCAS and FORS2, 9 out of 18 candidates
(50\%) are spectroscopically identified as SN~Ia or SN~Ia*. Of these
9, 8 are below $z=1.1$. FORS2 and FOCAS proved to be equally
effective in this redshift interval.

Clearly, $z>1.1$ SNe~Ia are more difficult to confirm
spectroscopically with ground based facilities. This is not a
surprising result. They are fainter and the peak of the flux is
shifted into a wavelength region where the night sky is
bright. However, it is possible to confirm SNe beyond $z=1.1$ with
optical instrumentation, if conditions are ideal and integrations are
long. SCP06G4 in this paper is an example. With the upgrade of optical
spectrographs with CCDs that have improved red sensitivity and minimal
fringing (as was recently done for LRIS on Keck and is now being 
done for FOCAS on Subaru), 
it should be possible to increase the yield of spectroscopically
classified SNe~Ia with ground based facilities beyond $z=1.1$. 
Laser guide star adaptive optics
(LGS-AO) assisted infra-red spectroscopy may be another alternative;
however, only a few SNe~Ia are expected to have a tip-tilt star that
is both near enough and bright enough for LGS-AO to be
effective. SCP05D6, for example, was imaged in the $H$-band
(rest frame $R$-band) with the LGS-AO system at the Keck Observatory
\citep{melbourne2007}.

The strategy adopted in the HST Cluster SN Survey is an
alternative. Since Type II SN are extremely rare in early-type
galaxies, spectroscopic confirmation of the SN type in such
galaxies is unnecessary. 
Spectroscopy is still necessary to obtain
redshifts. However, the amount of time needed to obtain redshifts is
generally much less than the time needed to confirm the SN type, and
the spectrum can be taken long after the SN is no longer is
visible. In one case, SCP05D0, a spectrum of the host was
  available from archived data and was taken before the SN 
exploded \citep{dawson2009}.
The key here is to show that the amount of star formation in the host
is low enough that SN of other types are unlikely.

\section{Summary}\label{sec:summary}

We presented spectra of SN candidates obtained with FOCAS on the
Subaru 8.2-m telescope. Seven active candidates were identified 
as SNe Ia, including SCP06G4 at $z_{\rm{SN}}=1.35$, 
the most distant SN~Ia to be spectroscopically identified 
with a ground-based telescope. 
Redshifts were obtained for all but 7 of the remaining 32 candidates 
based on the host galaxy spectra.
An additional 4 candidates are identified as likely SNe~Ia from the
spectrophotometric properties of their hosts.  The spectral properties
of these hosts found in the HST Cluster Supernova Survey were examined
by comparing their spectra with BC03 model spectra. All of the host
galaxy spectra indicate that they are passively evolving galaxies that
have quenched their star forming activities.

We also investigated the factors affecting the classification of SNe
and found that it is critical to take spectra within a week of maximum light. This requires 
secure early detection, well-sampled light curves and prompt spectroscopic follow-up.

\bigskip

We thank the anonymous referee for providing helpful comments
and suggestions. TM and YI are financially supported by the Japan
Society for the Promotion of Science (JSPS) through the JSPS Research
Fellowship.  
CL acknowledges the financial support from the Oskar Klein Centre 
at the University of Stockholm. 
This work was supported in part with scientific research
grants (15204012 and 17104002) from the Ministry of Education,
Science, Culture, and Sports of Japan, and a JSPS core-to-core program
``International Research Network for Dark Energy''. Financial support
for this work was provided in part by NASA through program GO-10496
from the Space Telescope Science Institute, which is operated by AURA,
Inc., under NASA contract NAS 5-26555.  This work was also supported
in part by the Director, Office of Science, Office of High Energy and
Nuclear Physics, of the U.S. Department of Energy under Contract
No. AC02-05CH11231.  Part of the Suprime-Cam observations were done
during the guaranteed time observation of Suprime-Cam and we thank for
the Suprime-Cam instrument team.  We also appreciate much help by the
SDF and SXDS project team members.  We also thank for Youichi Ohyama,
who helped our observations as a support scientist of FOCAS.  Data
analysis were in part carried out on common use data analysis computer
system at the Astronomy Data Center, ADC, of the National Astronomical
Observatory of Japan.


\end{document}